\documentclass[twocolumn, eqsecnum, prd, superscriptaddress, nofootinbib]{revtex4-2}

\usepackage{amsmath,amssymb}
\usepackage{amsthm}
\usepackage{mathtools}
\usepackage{appendix}
\usepackage{comment}
\usepackage{graphicx}
\usepackage{grffile}
\usepackage{mathrsfs}
\usepackage{bm}
\usepackage{url}
\usepackage[colorlinks=true
,urlcolor=DARKBLUE
,anchorcolor=DARKBLUE
,citecolor=DARKBLUE
,filecolor=DARKBLUE
,linkcolor=DARKBLUE
,menucolor=DARKBLUE
,linktocpage=true
,pdfproducer=medialab
,pdfa=true
]{hyperref}
\usepackage{xcolor}
\usepackage{soul}
\usepackage{algorithm}
\usepackage{algorithmic}
\usepackage{natbib}
\usepackage{physics}
\usepackage{siunitx}
\usepackage{acronym}
\usepackage{xspace}

\newcommand{\diff}[2]{\dv{#1}{#2}}

\newcommand{\fpbh}{f_\textrm{PBH}}
\newcommand{\mpbh}{M_\textrm{PBH}}

\newcommand{\Rpbh}{R_\mathrm{PBH}}
\newcommand{\mwd}{M_\textrm{WD}}

\newcommand{\Rwd}{R_\mathrm{WD}}

\newcommand{\PBH}{\mathrm{PBH}}
\newcommand{\WD}{\mathrm{WD}}

\newcommand{\Kepler}{\mathrm{Kepler}}
\newcommand{\obs}{\mathrm{obs}}

\newcommand{\tidal}{\mathrm{tidal}}
\newcommand{\halo}{\mathrm{halo}}
\newcommand{\MS}{\mathrm{MS}}
\newcommand{\Mc}{\mathcal{M}_\textrm{c}}

\newcommand{\df}{\dd{f}}

\newcommand{\fmin}{f_\mathrm{min}}
\newcommand{\fmax}{f_\mathrm{max}}
\newcommand{\fmergewb}{f_\mathrm{merge}^\text{WD-PBH}}

\newcommand{\Hz}{\mathrm{Hz}}
\newcommand{\etal}{\textit{et al.}\xspace}
\newcommand{\vrel}{v_\mathrm{rel}}
\newcommand{\vdm}{v_\mathrm{DM}}
\newcommand{\vvir}{v_\mathrm{vir}}
\newcommand{\umin}{\mathrm{min}}
\newcommand{\umax}{\mathrm{max}}

\definecolor{MONZA}{HTML}{CF000F}
\definecolor{DARKBLUE}{HTML}{00008b}
\definecolor{DARKMAGENTA}{HTML}{8b008b}
\definecolor{DARKCYAN}{HTML}{00cfc0}
\definecolor{brightpink}{rgb}{1.0, 0.0, 0.5}

\newcommand{\uh}{\mathrm{h}}
\newcommand{\um}{\mathrm{m}}
\newcommand{\calO}{\mathcal{O}}
\newcommand{\us}{\mathrm{s}}

\newcommand{\ee}{\mathrm{e}}
\newcommand{\rel}{\mathrm{rel}}

\newcommand{\beae}[1]{\begin{equation}\begin{aligned} #1 \end{aligned}\end{equation}}

\newcommand{\bme}[1]{\begin{multline} #1 \end{multline}}
\newcommand{\bmte}[1]{\begin{multlined}[t] #1 \end{multlined}}

\begin{document}
\title{Prospects of detection of subsolar mass primordial black hole and white dwarf binary mergers}
\author{Takahiro S. Yamamoto}
\email{yamamoto.takahiro.u6@f.mail.nagoya-u.ac.jp}
\author{Ryoto Inui}
\email{inui.ryoto.a3@s.mail.nagoya-u.ac.jp}
\affiliation{Department of Physics, Nagoya University, 
Furo-cho Chikusa-ku,
Nagoya 464-8602, Japan}
\author{Yuichiro Tada}
\email{tada.yuichiro.y8@f.mail.nagoya-u.ac.jp}
\affiliation{Institute for Advanced Research, Nagoya University,
Furo-cho Chikusa-ku, 
Nagoya 464-8601, Japan}
\affiliation{Department of Physics, Nagoya University, 
Furo-cho Chikusa-ku,
Nagoya 464-8602, Japan}
\author{Shuichiro Yokoyama}
\email{shu@kmi.nagoya-u.ac.jp}
\affiliation{Kobayashi Maskawa Institute, Nagoya University, 
Chikusa, Aichi 464-8602, Japan}
\affiliation{Kavli IPMU (WPI), UTIAS, The University of Tokyo, 
Kashiwa, Chiba 277-8583, Japan}

\date{\today}
\begin{abstract} 
The subsolar mass \ac{PBH} attracts attention as robust evidence of its primordial origin against the astrophysical black hole.
Not only with themselves, \acp{PBH} can also form binaries with ordinary astrophysical objects, catching them by \ac{GW} bremsstrahlung.
We discuss the detectability of the inspiral \acp{GW} from binaries consisting of a \ac{PBH} and a \ac{WD} by using space-borne gravitational wave interferometers like DECIGO. The conservative assessment shows the expected event number in three years by DECIGO is $\calO(10^{-6})$ for $\mpbh \sim 0.1M_\odot$.
Possible enhancement mechanisms of \ac{WD}-\ac{PBH} binary formation may amplify this event rate. We discuss how large enhancement associated with \acp{WD} is required to detect WD-PBH merger events without violating the existing constraints on the \ac{PBH}-\ac{PBH} merger by the ground-based detector.
\end{abstract}
\maketitle
\acresetall

\acrodef{SNR}{signal-to-noise ratio}
\acrodef{PSD}{power spectral density}
\acrodef{PBH}{primordial black hole}
\acrodef{NS}{neutron star}
\acrodef{WD}{white dwarf}
\acrodef{GW}{gravitational wave}
\acrodef{NFW}{Navarro--Frenk--White}
\acrodef{LVK}{LIGO--Virgo--KAGRA}

\section{introduction}
\label{sec: introduction}
\Acp{PBH} are hypothetical black holes that could have formed in the early universe before star formation~\cite{Zeldovich:1967lct, Hawking1971, Carr:1974nx, Carr:1975qj}. 
They have attracted attention more and more as a candidate for dark matter~\cite{Chapline:1975ojl}, seeds of supermassive black holes~\cite{Bean:2002kx, Duechting:2004dk} (see, e.g., Refs.~\cite{Liu:2022bvr,Hutsi:2022fzw} for implications of supermassive \ac{PBH} by high-redshift luminous galaxies found by James Webb Space Telescope), planetary-mass microlensing objects towards the Galactic bulge detected by the Optical Gravitational Lensing Experiment~\cite{Niikura:2019kqi}, hypothetical Planet 9 in our solar system~\cite{Scholtz:2019csj,Witten:2020ifl}, triggers of faint supernovae called Calcium-rich gap transients~\cite{Smirnov:2022zip}, etc. (see, e.g., Refs.~\cite{Escriva:2022duf,Carr:2023tpt} for recent reviews on astrophysical implications of \acp{PBH}).
A significant property of \acp{PBH} is that they can have a wide range of mass even smaller than the solar mass, unlike the astrophysical black holes that are formed through the gravitational collapse of massive stars. Thus, the detection of subsolar mass black holes is a smoking gun for the existence of \Acp{PBH}.

While subsolar mass \acp{PBH} have been searched mainly by the gravitational microlensing\footnote{So far any microlensing search for \acp{PBH} has not detected a \ac{PBH}, but several candidate events are reported, represented by the single event by the Subaru Hyper Suprime-Camera~\cite{Niikura:2017zjd}.} so far (see Ref.~\cite{Carr:2020gox} for a summary of the existing observational constraints on the \ac{PBH} abundance), \acp{GW} have been attracting much attention as a new tool to probe the subsolar mass \Acp{PBH}. In the early universe, \Acp{PBH} can form binaries by the tidal torque exerted by a neighboring PBH~\cite{Nakamura:1997sm, Ioka:1998nz, Ali-Haimoud:2017rtz}. After forming the binaries, their orbits gradually shrink as they emit the \Acp{GW}. Depending on the separation and the component masses of binaries, their \acp{GW} can have the frequency to which the \ac{GW} detectors are sensitive. \ac{LVK} collaboration has already obtained the constraints on the \ac{PBH} abundance from the search of the subsolar mass black hole binaries~\cite{LIGOScientific:2018glc, LIGOScientific:2019kan, LIGOScientific:2021job, LIGOScientific:2022hai}. Several groups independently report candidates of \Acp{GW} from subsolar mass black hole binaries~\cite{Phukon:2021cus, Morras:2023jvb, Prunier:2023cyv}. Furthermore, Pujolas~\etal~\cite{Pujolas:2021yaw} estimated the detectability of the \ac{PBH} binaries in various future observations, such as Einstein Telescope~\cite{Maggiore:2019uih}, LISA~\cite{LISA:2017pwj}, and DECIGO~\cite{Seto:2001qf, Kawamura:2020pcg}.

While the dominant formation channel of the \ac{PBH} binaries is the three-body interaction in the early universe, the \ac{GW} bremsstrahlung can lead to \ac{PBH} binaries in the late universe~\cite{QuinlanShapiro1989, Mouri:2002mc, Ali-Haimoud:2017rtz}. In this channel, an unbounded \ac{PBH} is captured by another \ac{PBH} through the close encounter due to the energy loss by the \ac{GW} emission. This channel can form not only PBH-PBH binaries but also the binaries of a \ac{PBH} and another astrophysical compact object. For example, Tsai~\etal~\cite{Tsai:2020hpi} and Sasaki~\etal~\cite{Sasaki:2021iuc} estimate the merger rate of binaries consisting of a \ac{PBH} and a \ac{NS}. Tsai~\etal discussed the scenario in which GW170817 is an NS-PBH merger event. Sasaki~\etal shows that the NS-PBH binaries are subdominant, and their merger rate is not enough to explain \ac{LVK}'s NS-BH merger candidates.

In this work, as another system, we consider the binaries consisting of a subsolar mass \ac{PBH} and a \ac{WD}. The merger frequencies of the WD-PBH binaries are $\mathcal{O}(10^{-2} \text{--} 10^{-1})\,\si{Hz}$, which are interesting targets of DECIGO. 
However, numerical simulations of the \ac{WD}-\ac{PBH} merger are yet to be done.
We then focus on the inspiral \ac{GW} and investigate the detectability of the WD-PBH merger event via the maximum detectable distance of inspiral events with a sufficiently large \ac{SNR} in the DECIGO observation. 
As \ac{WD}-\ac{PBH} inspiral \acp{GW} can be mimicked by \ac{WD}-\ac{WD} ones, we also discuss whether they can be distinguished only by the differences in the merger frequencies.
Finally, we estimate the expected event number of the WD-PBH binaries in the three-year DECIGO mission.

The paper is organized as follows.
In Sec.~\ref{sec: detectability}, we discuss the detectability and the distinguishability of WD-PBH binaries by assuming three years of observation of DECIGO. In Sec.~\ref{sec: expected number of events}, we present the merger rate estimation, and Sec.~\ref{sec: conclusion} is devoted to the conclusions and future perspective. The details of the merger rate calculation are summarized in Appendix.~\ref{appendix}.
Throughout the paper, we suppose that the \ac{PBH} mass function is monochromatic.

\section{Detectability by DECIGO}
\label{sec: detectability}

\subsection{Signal-to-noise ratio for the inspiraling WD-PBH binary}

\begin{figure}
    \centering
    \includegraphics[width=8.5cm]{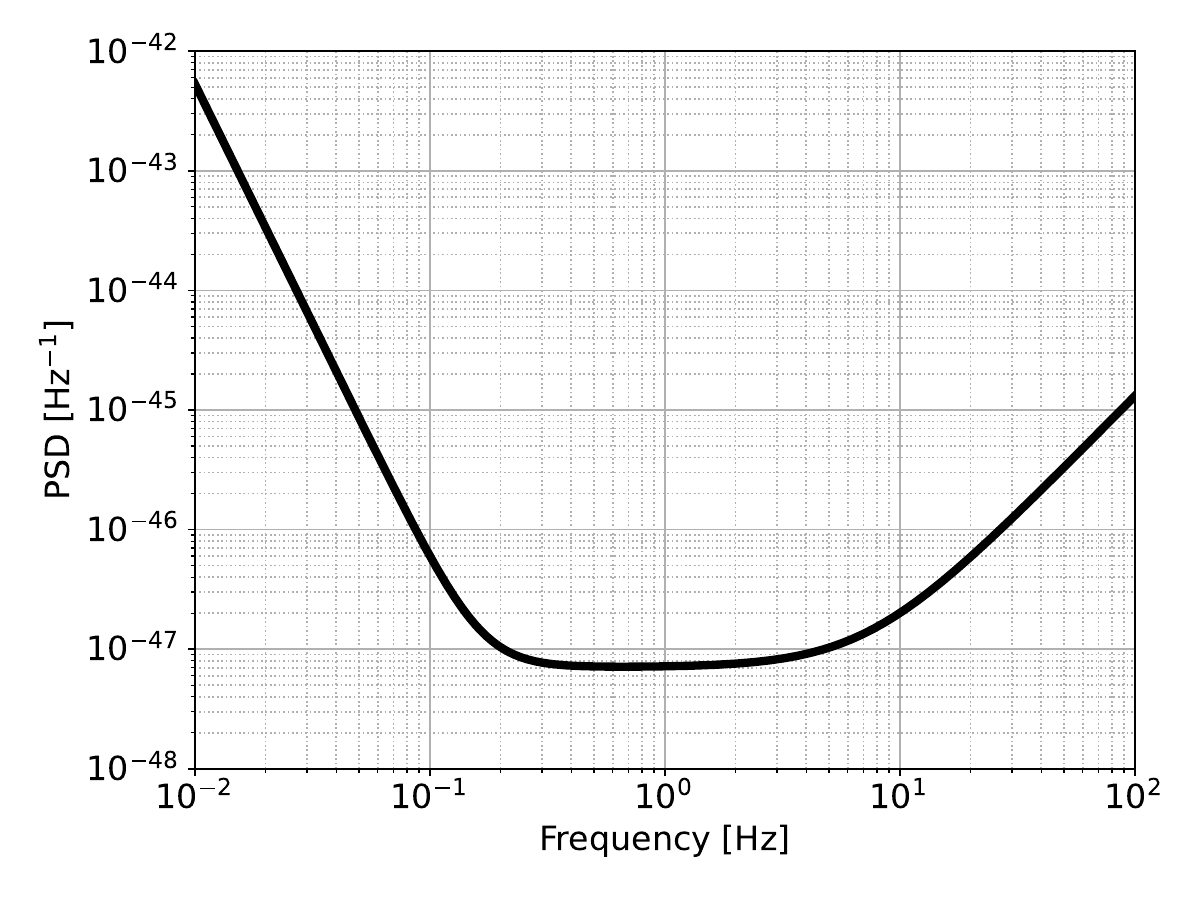}
    \caption{Noise PSD of DECIGO~\eqref{eq: Sn}.}
    \label{fig:decigo psd}
\end{figure}

In this subsection, we evaluate the detectability of WD-PBH binaries only through the inspiral \acp{GW}. The most optimistic scenario, which we hereafter assume, is that DECIGO keeps observing the binary inspiral for three years from the beginning of the observation run, and the binary merges just after the end of the observation run. The detectability is quantified by the \ac{SNR} of the inspiral waveform from the initial frequency $\fmin$ to the final frequency $\fmax$ in the detector frame.
Note that we basically suppose quasi-circular orbits for \ac{WD}-\ac{PBH} binaries for brevity though Cholis~\etal~\cite{Cholis:2016kqi} suggested that the binaries formed in the \ac{GW} bremsstrahlung have high eccentricities and merge within a few years, typically keeping their eccentricities even at the merger.

The measured signal in the frequency domain is obtained by
\begin{equation}
    \tilde{h}(f) = F_+ \tilde{h}_+(f) + F_\times \tilde{h}_\times(f)\,,
\end{equation}
where $F_+$ and $F_\times$ are the antenna pattern functions, and $\tilde{h}_{+/ \times}$ are plus/cross modes expressed by
\begin{equation}
    \tilde{h}_+(f) = (1 + \cos^2\iota) \tilde{h}_0(f)\,,\quad
    \tilde{h}_\times(F) = -2 i \cos\iota \tilde{h}_0(f)\,,
\end{equation}
with the inclination angle $\iota$.
$\tilde{h}_0$ is given by
\begin{equation}
   \tilde{h}_0(f) = \sqrt{\frac{5}{96}} \pi^{-2/3} \qty(\frac{(1+z)G\Mc}{c^3})^{5/6} \frac{c}{d_\mathrm{L}} f^{-7/6} \ee^{i\Psi(f)}\,,
\end{equation}
where $\Psi(f)$ is the \ac{GW} phase, $G$ is the Newtonian constant of gravity, and $c$ is the speed of light. $\Mc$ is the chirp mass defined by
\begin{equation}
    \Mc \coloneqq \frac{(m_1 m_2)^{3/5}}{(m_1 + m_2)^{1/5}}\,,
\end{equation}
with the component masses $m_1$ and $m_2$ in the source frame. 
$z$ and $d_\mathrm{L}$ are the source redshift and the luminosity distance, respectively. Their relation is given by
\begin{equation}
    d_\textrm{L} = (1+z) \int^z_0 \frac{c\dd{z}}{H(z)}\,.
\end{equation}
We employ \texttt{astropy}~\cite{astropy} to calculate the luminosity distance and use the cosmological parameters obtained by Planck18~\cite{Planck:2018vyg}.
Given the detector noise spectrum $S_\mathrm{n}(f)$, we define the \ac{SNR} by
\begin{equation}
    \varrho_0 = \qty[ 4\int^\infty_0 \frac{|\tilde{h}(f)|^2}{S_\mathrm{n}(f)} \dd{f}]^{1/2}\,.
\end{equation}
Following Ref.~\cite{Dalal:2006qt}, we estimate the \ac{SNR} averaged over the extrinsic parameters (e.g., the source position in the sky) by the formula
\begin{equation}
    \varrho = \frac{8}{5}\frac{A}{d_\textrm{L}} \sqrt{I_{7}}\,,
    \label{eq: inspiral signal 1}
\end{equation}
where $I_7$ and $A$ are given by
\begin{align}
    &I_{7} = \int^{\fmax}_{\fmin} \df\ \frac{f^{-7/3}}{S_\textrm{n}(f)}\,,
    \label{eq:inner product} \\
    &A = \sqrt{\frac{5}{96}} \pi^{-2/3} c \left( \frac{(1+z) G\Mc}{c^3} \right)^{5/6}\,.
    \label{eq: inspiral signal 2}
\end{align}
The integration range $[\fmin, \fmax]$ in Eq.~\eqref{eq:inner product} is obtained below as a frequency evolution during the three-year observation of DECIGO.

In the present work, we use the DECIGO's design \ac{PSD}~\cite{Yagi:2011wg}
\begin{widetext}
    \begin{equation}\label{eq: Sn}
        S^\mathrm{DECIGO}_\textrm{n}(f) = \bqty{7.05 \times 10^{-48} \qty( 1 + \left( \frac{f}{f_\textrm{p}} \right)^2)
        + 4.8 \times 10^{-51}\qty(\frac{f}{\SI{1}{Hz}})^{-4} \frac{1}{1 + \qty(\frac{f}{f_\textrm{p}})^2 }
        + 5.33 \times 10^{-52} \qty(\frac{f}{\SI{1}{Hz}})^{-4}}\,\si{Hz^{-1}},
    \end{equation}
\end{widetext}
with $f_\textrm{p} = \SI{7.36}{Hz}$.
Figure~\ref{fig:decigo psd} shows the noise \ac{PSD} of DECIGO.

Here, we focus only on \Acp{GW} from the inspiral phase.
We assume that $\fmax$ is 
given by
\begin{align}
f_\textrm{max} = \frac{f_\textrm{merge}^\textrm{WD-PBH}}{1 + z}\,,
\end{align}
where $\fmergewb$ is a typical merger frequency in the source frame.
In the case of WD-PBH binaries, \Acp{WD} would be tidally disrupted in the final stage of the inspiral phase, and then
$\fmergewb$ is determined by the binary separation equal to the tidal radius, $R_\textrm{tidal}$, which is estimated by balancing the tidal force and the WD's self-gravity as
\begin{align}
    \label{eq:tidal}
    &F_\textrm{tidal} = F_\textrm{grav}
    \quad\text{at $r=R_\tidal$}\notag\\
    \Rightarrow\quad 
    &\frac{2G \mpbh}{r^2} \left( \frac{\Rwd}{r} \right) = \frac{G\mwd}{\Rwd^2} 
    \quad\text{at $r=R_\tidal$}\notag\\
    \Rightarrow\quad
    &R_\tidal = \left(\frac{2\mpbh}{\mwd} \right)^{1/3} \Rwd\,,
\end{align}
where $\mwd$ and $\mpbh$ are respectively masses of the \ac{WD} and the PBH, and $\Rwd$ is the radius of the \ac{WD}.
We use the following mass-radius relation:
\begin{equation}
    R_\textrm{WD} = 0.0126R_\odot \left(\frac{\mwd}{M_\odot}\right)^{-1/3}\,,
\end{equation}
which is proposed in Ref.~\cite{Magano:2017mqk}.
The corresponding frequency is given by 
\begin{equation}
    f_\textrm{tidal} = \frac{1}{\pi} \sqrt{\frac{G (\mwd + \mpbh)}{R_\textrm{tidal}^3}}\,.
\end{equation}

Equation~\eqref{eq:tidal} implies that $R_\textrm{tidal}$ is smaller than the radius of the \ac{WD}, $\Rwd$, if $2\mpbh < \mwd$.
In such cases, as the merger frequency, $\fmergewb$, we should use the Kepler frequency at the time when the orbital radius becomes the sum of the radii of the \ac{WD} and the \ac{PBH} ($\Rpbh$):
\begin{equation}
    f_\textrm{Kepler} = \frac{1}{\pi} \sqrt{ \frac{G (\mwd + \mpbh)}{(\Rwd + \Rpbh)^3} }.
\end{equation}
The merger frequency is determined by the one realized earlier:
\begin{equation}\label{eq: f merger pbh-wd}
    \fmergewb = \min[f_\textrm{Kepler}, f_\textrm{tidal}]\,.
\end{equation}

As for $\fmin$, we introduce $f_\textrm{start}$ as the corresponding frequency in the source frame:
\begin{equation}
f_\textrm{min} = \frac{f_\textrm{start}}{1 + z}\,.
\end{equation}
The duration that the binary requires to evolve from the frequency $f_1$ to $f_2$ ($>f_1$) is given by (see Refs.~\cite{Creighton:2011zz, Maggiore:2007ulw, Jaranowski:2009zz})
\begin{equation}
    t(f_2) - t(f_1) = \frac{5}{256} \left( \frac{G \Mc}{c^3} \right)^{-5/3} \pi^{-8/3} (f_1^{-8/3} - f_2^{-8/3})\,.
\end{equation}
From this expression, by assuming that the binary merges at the end of the observation period in our optimistic scenario,
we can evaluate the frequency at the onset of the observation run (i.e., three years ($=T_\obs$) before the merger),
which is corresponding to $f_\textrm{start}$, as
\begin{equation}
T_\textrm{obs} = t(\fmergewb) - t(f_\textrm{start})\,,
\end{equation}
and it gives
\begin{align}
   &f_\textrm{start} \nonumber \\
   &= \left\{ \qty(f_\textrm{merge}^\textrm{WD-PBH})^{-8/3} + T_\textrm{obs} \frac{256 \pi^{8/3}}{5} \left( \frac{G \Mc}{c^3} \right)^{5/3} \right\}^{-3/8}. 
\end{align}

\begin{figure}
    \centering
    \includegraphics[width=
    0.9\hsize]{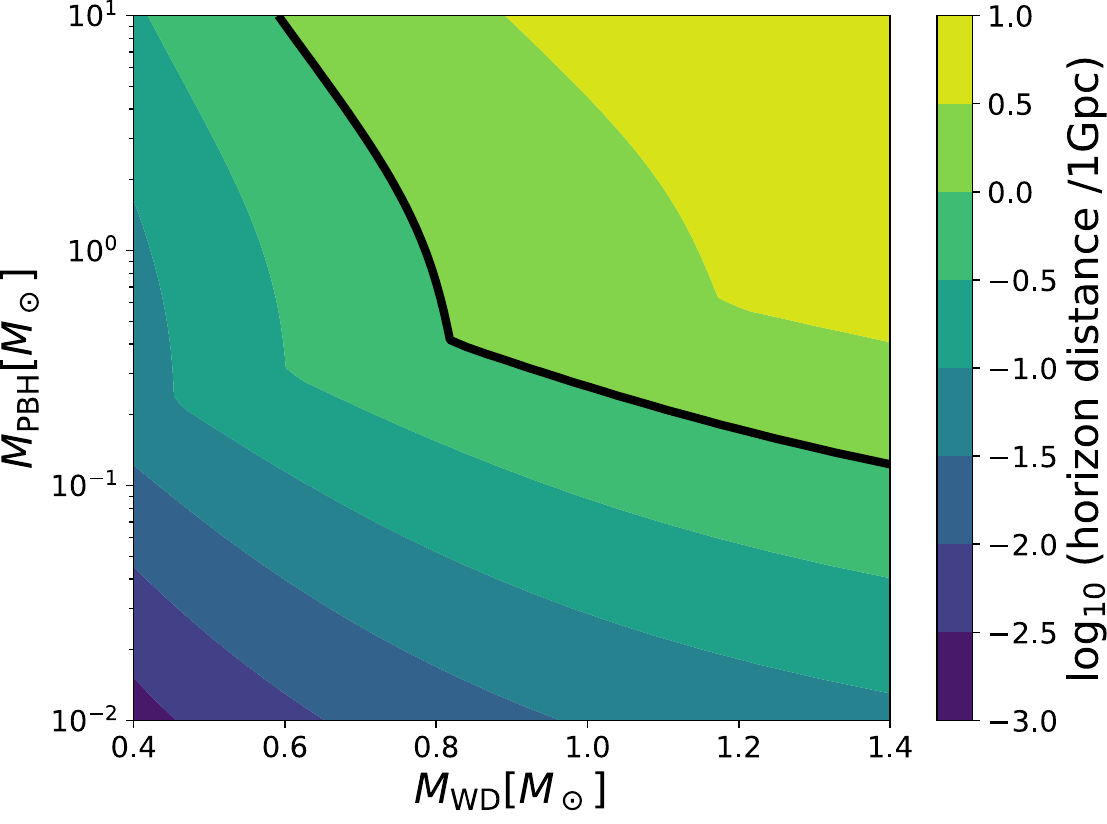}
    \caption{Horizon distance as a function of $M_\textrm{PBH}$ and $M_\textrm{WD}$. The black solid line indicates $\SI{1}{Gpc}$. There are breaks around $\mpbh \sim (2\text{--}5)\times 10^{-1}M_\odot$. Around there, the merger frequency, defined by Eq.~\eqref{eq: f merger pbh-wd}, changes drastically because the tidal frequency $f_\tidal$ and the Kepler frequency $f_\Kepler$ have different mass dependencies.}
    \label{fig:horizon distance}
\end{figure}

We set the detection threshold on the \ac{SNR} at $8$~\cite{Maggiore:2007ulw} and refer to the maximum distance of the detectable binaries as the horizon distance. Figure~\ref{fig:horizon distance} shows the horizon distance depending on the masses of a \ac{PBH} and a \ac{WD}. From this figure, we can find that we would detect the \Acp{GW} from a \ac{WD}-\ac{PBH} binary at $\sim\SI{1}{Gpc}$ for $M_\textrm{WD} \gtrsim 0.8M_\odot$ and $M_\textrm{PBH} \gtrsim 0.5M_\odot$.

\onecolumngrid

\begin{figure*}
     \begin{tabular}{ccc}
        \begin{minipage}{0.33\hsize}
            \centering
            \includegraphics[width=5.9cm]{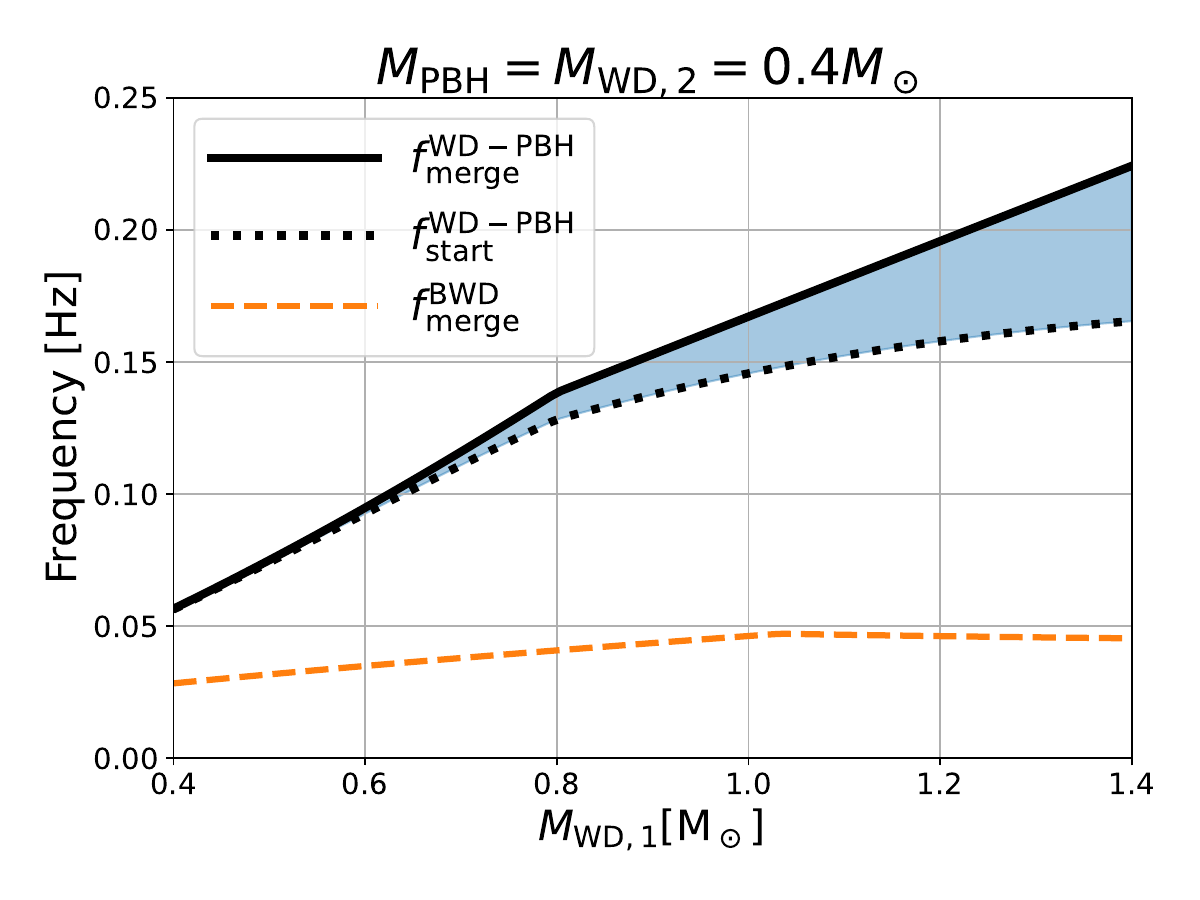}
        \end{minipage} &
        \begin{minipage}{0.33\hsize}
            \centering
            \includegraphics[width=5.9cm]{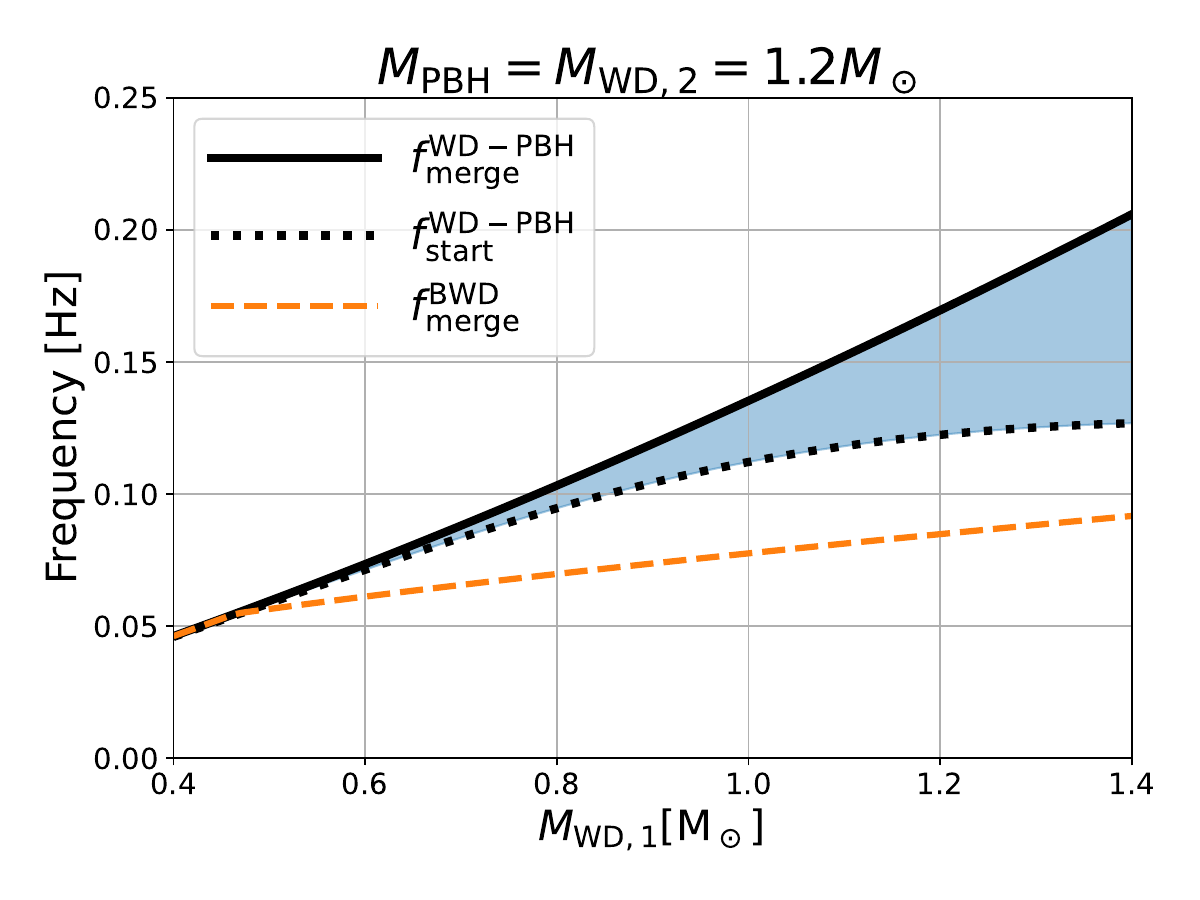}
        \end{minipage} &
        \begin{minipage}{0.33\hsize}
            \centering
            \includegraphics[width=5.9cm]{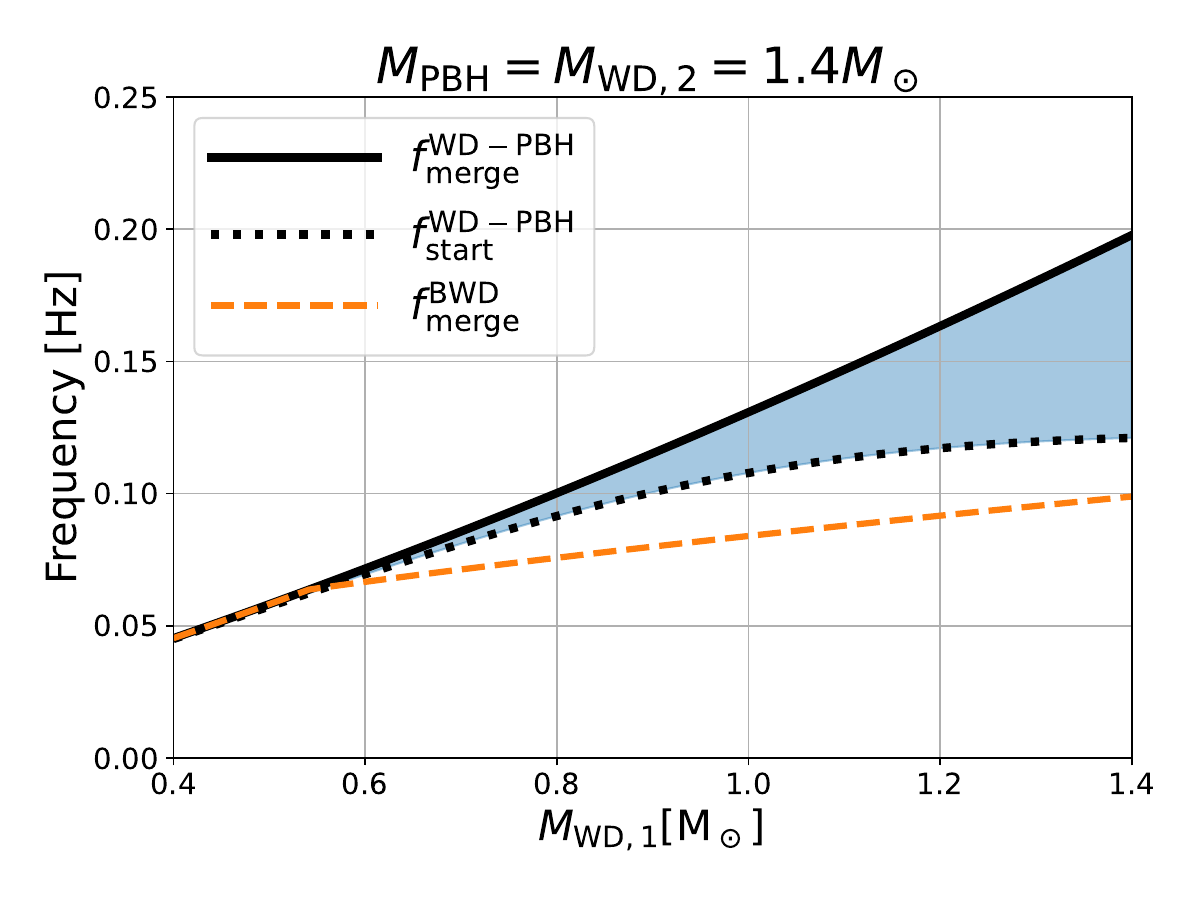}
        \end{minipage}
    \end{tabular}
    \caption{Comparison of the relevant frequencies for \ac{WD}-\ac{PBH} and \ac{WD}-\ac{WD} binaries. Black solid and dotted lines are maximum and minimum frequencies realized by \ac{WD}-\ac{PBH} inspirals, respectively. Orange dashed lines show the merger frequency of the WD-WD binary whose component masses are $M_\textrm{WD,1}$ and $M_\textrm{WD,2}$. If the orange dashed line is below the black dotted line, \ac{WD}-\ac{PBH} and \ac{WD}-\ac{WD} binaries can be distinguished only by the inspiral waveform.
    We assume three years of observation.}
    \label{fig: merger frequency}
\end{figure*}
\twocolumngrid

\subsection{Distinguishability between WD-PBH and WD-WD}
In the previous subsection, we evaluated the detectability of WD-PBH binaries.
The \ac{GW} waveform from a compact binary is characterized by the chirp mass $\Mc$. Therefore, if a WD-WD binary and a WD-PBH binary have similar component masses, they can not be distinguished only from the chirp mass. However, the merger frequency can be different because of the difference in the radii of \Acp{WD} and \Acp{PBH}. We discuss, here, how we can distinguish WD-WD binaries and WD-PBH binaries by the difference in the merger frequencies.

The merger frequency of WD-WD binaries, which is denoted by $f_\textrm{merge}^\textrm{BWD}$, is estimated as well as the merger frequency of WD-PBH binaries. We consider the WD-WD binary with the masses of $M_\textrm{WD,1}$ and $M_\textrm{WD,2}$. We use $R_\textrm{tidal,1}$ to denote the tidal radius in which the first \ac{WD} ($M_\textrm{WD,1}$ and $R_\textrm{WD,1}$) is disrupted by the companion \ac{WD} ($M_\textrm{WD,2}$). It is estimated by
\begin{equation}
    R_\textrm{tidal,1} = \left( \frac{2M_\textrm{WD,2}}{M_\textrm{WD,1}} \right)^{1/3} R_\textrm{WD,1}\,.
\end{equation}
By changing the labels $1 \leftrightarrow 2$, the tidal radius $R_\textrm{tidal,2}$ can be defined.
When the two \acp{WD} approach as close as the radius
\begin{equation}
    R_\textrm{tidal}^\textrm{BWD} = \max[R_\textrm{tidal,1}, R_\textrm{tidal,2}]\,,
\end{equation}
the lighter one is tidally disrupted.
We define the frequency corresponding to $R_\textrm{tidal}^\textrm{BWD}$ by
\begin{equation}
    f_\textrm{tidal}^\textrm{BWD} = \frac{1}{\pi} \sqrt{\frac{G (M_\mathrm{WD,1} + M_\mathrm{WD,2})}{\qty(R_\textrm{tidal}^\textrm{BWD})^3}}\,.
\end{equation}
When $M_\textrm{WD,1} \simeq M_\textrm{WD,2}$, $R_\textrm{tidal}^\textrm{BWD}$ could be smaller than the sum of the WD's radii. In such cases, as the merger frequency, we should employ the Kepler frequency at the time when the binary separation is equal to the sum of the WD's radii:
\begin{equation}
    f_\textrm{Kepler}^\textrm{BWD} = \frac{1}{\pi}\sqrt{\frac{G(M_\textrm{WD,1} + M_\textrm{WD,2})}{(R_\textrm{WD,1} + R_\textrm{WD,2})^3}}\,.
\end{equation}
Thus, as in the case of the WD-PBH binary, the merger frequency of the WD-WD binary is given by the lower one:
\begin{equation}
    f_\textrm{merge}^\textrm{BWD} = \min[f_\textrm{Kepler}^\textrm{BWD}, f_\textrm{tidal}^\textrm{BWD}]\,.
\end{equation}

Figure~\ref{fig: merger frequency} shows the comparisons between the merger frequencies, $f^\textrm{WD-PBH}_\textrm{merge}$ and $f^\textrm{BWD}_\textrm{merge}$, and the initial frequency $f^\textrm{WD-PBH}_\textrm{start}$ of the WD-PBH binary as functions of $\mpbh$. If the frequency satisfies $f_\textrm{merge}^\textrm{BWD} \leq f_\textrm{start}^\textrm{WD-PBH}$, the WD-PBH binary merges at the frequency that WD-WD binaries with the same masses never reach. So, we can distinguish the WD-PBH binary and the WD-WD binary by the difference in the merger frequencies.

\begin{figure}
    \centering
    \includegraphics[width=8cm]{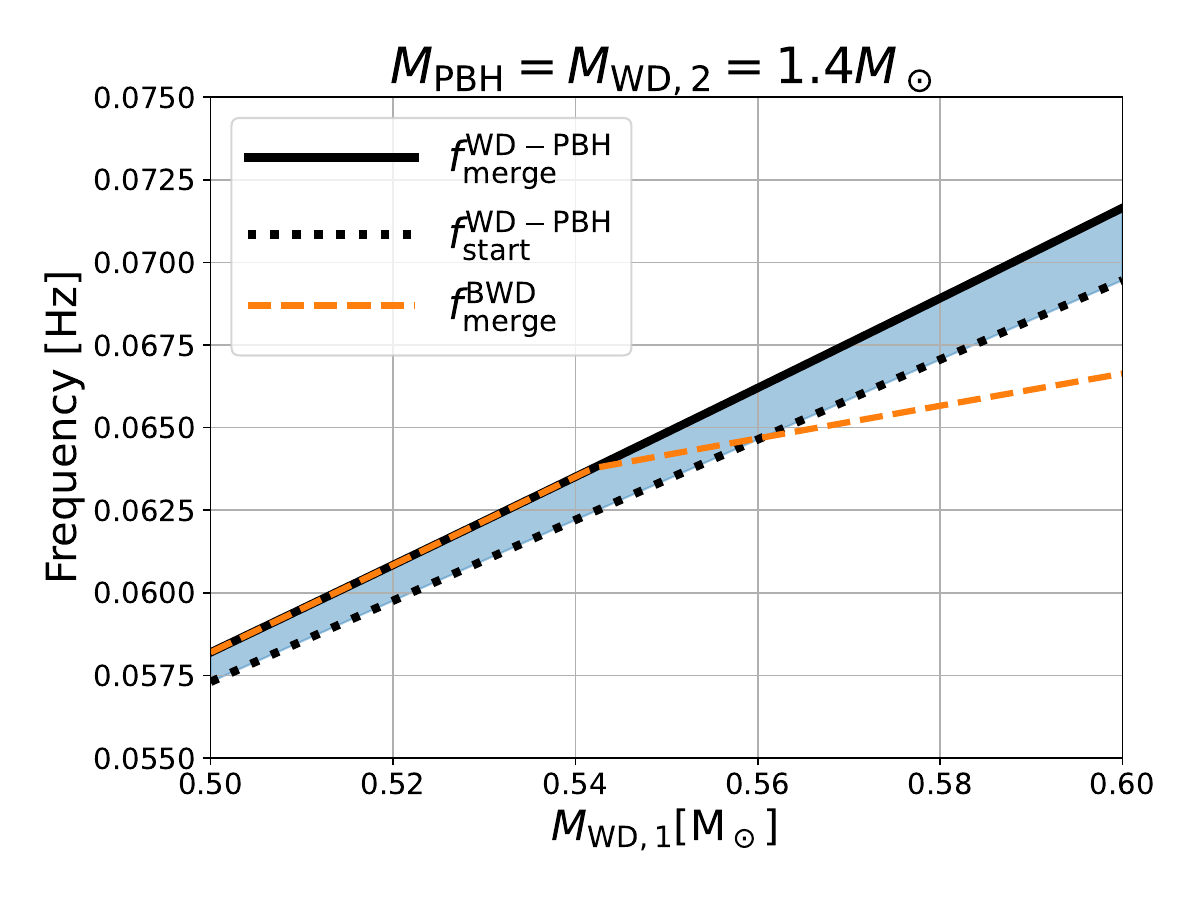}
    \caption{Enlarged figure of the right panel of Fig.~\ref{fig: merger frequency}. While binaries are distinguishable for $M_{\WD,1}\gtrsim0.56M_\odot$, it is hard for $M_{\WD,1}\lesssim0.56M_\odot$ and impossible for $M_{\WD,1}\lesssim0.54M_\odot$ to distinguish them only by the inspiral waveform and hence the merger waveform is required which is beyond the scope of the paper.}
    \label{fig: merger frequency zoom}
\end{figure}

\begin{figure}
    \centering
    \includegraphics[width=8cm]{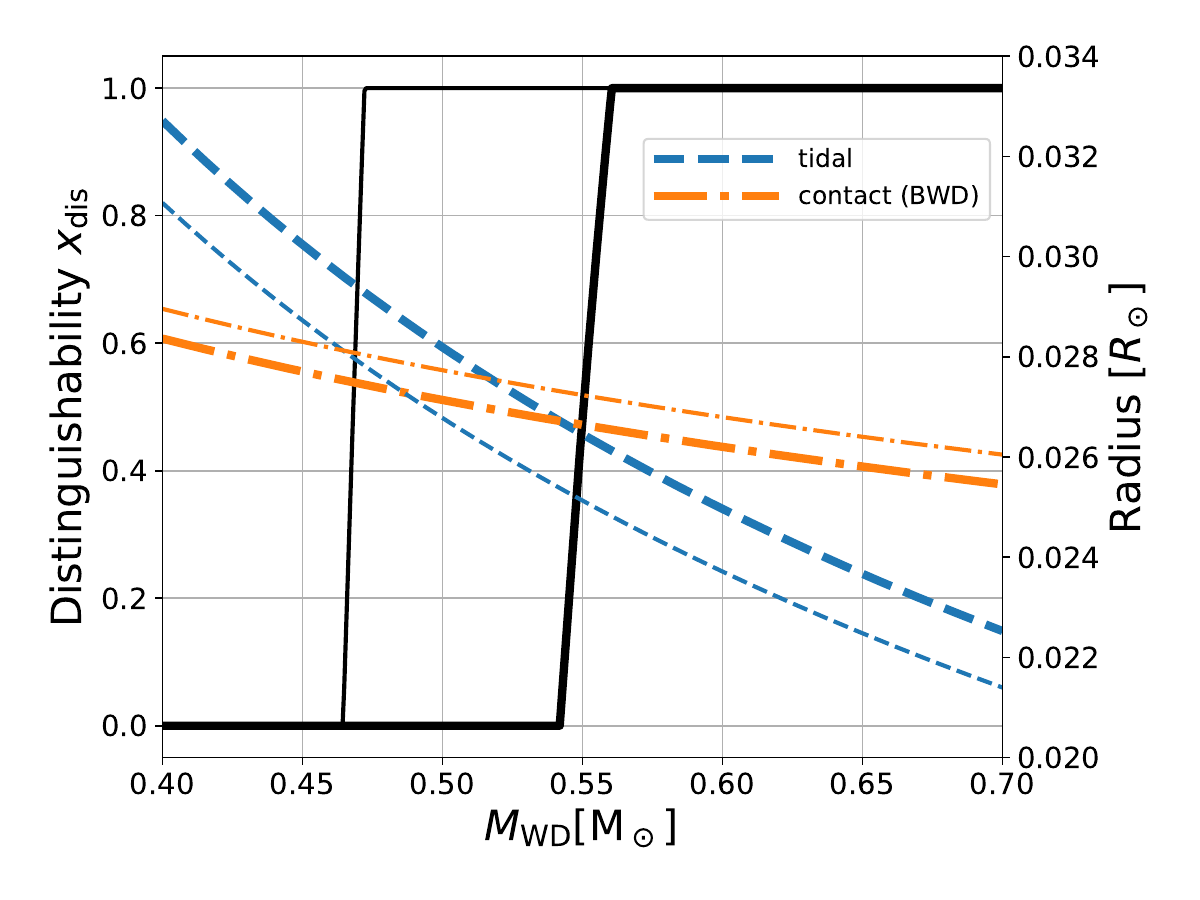}
    \caption{Distinguishability~\eqref{eq: overlap} in black lines. The thick (thin) line shows the \ac{PBH} mass of 1.4 $M_\odot$ (1.2 $M_\odot$). The distance to the source is chosen such that the reference \ac{SNR} $\varrho=30$. Although the integrals $I_7$ and $I'_7$ depend on the redshift, the result is not much affected by the value of the reference SNR.
    The blue dashed lines represent the tidal radius of the WD-PBH binary.
    The orange dash-dotted lines represent the contact radius of the WD-WD binary whose component masses are $\mwd$ and $1.4M_\odot$ ($1.2M_\odot$).}
    \label{fig: overlap}
\end{figure}

Figure~\ref{fig: merger frequency zoom} zooms in on the right panel of Fig.~\ref{fig: merger frequency}, focusing on the regions where $0.5~ M_\odot \lesssim M_{\WD,1} \lesssim 0.6 ~M_\odot$.
For some mass ranges,
the merger frequency of the WD-WD binary is located between $f^\textrm{WD-PBH}_\textrm{start}$ and $f^\textrm{WD-PBH}_\textrm{merge}$. 
This means that the waveforms of the WD-PBH binary and the WD-WD binary overlap in the frequency domain. If this overlap is significant, it might be difficult to distinguish between the WD-PBH and WD-WD binaries.
We define the quantity named ``distinguishability'' by
\begin{equation}
    x_\mathrm{dis} \coloneqq \frac{\varrho^2 - \varrho^{\prime 2}}{\varrho^2}\,,
    \label{eq: overlap}
\end{equation}
where $\varrho'$ is defined by
\begin{equation}
    \varrho' = \frac{8}{5} \frac{A}{d_\textrm{L}} \sqrt{I'_7}\,,
\end{equation}
with ``wrongly" replacing the \ac{WD}-\ac{PBH} merger frequency with that of the \ac{WD}-\ac{WD} one:
\begin{equation}
    I'_{7} = \begin{cases}
    \displaystyle
    \int^{f_\textrm{merge}^\textrm{BWD} / (1+z)}_{f_\textrm{start}^\textrm{WD-PBH}/(1+z)} \df 
    \frac{f^{-7/3}}{S_\textrm{n}(f)} & \text{for $f_\textrm{start}^\textrm{WD-PBH} < f_\textrm{merge}^\textrm{BWD}$,} \\
    0 & \text{for $f_\textrm{start}^\textrm{WD-PBH} \geq f_\textrm{merge}^\textrm{BWD}$.} 
    \end{cases}
\end{equation}
If $f^\textrm{BWD}_\textrm{merge}$ is equal or very close to $f^\textrm{WD-PBH}_\textrm{merge}$, we could not judge whether the binary is a WD-PBH binary or a WD-WD binary only from their inspiral parts because they have the same frequency evolution during the observational period. In such case, $\varrho'$ equals $\varrho$. In this sense, the ``distinguishability" $x_\mathrm{dis}$ is a quantity representing how likely we can distinguish a WD-PBH binary from a WD-WD binary.
That is, for $x_\mathrm{dis} \simeq 1$, we can distinguish a WD-PBH binary and a WD-WD binary, but for $x_\mathrm{dis} \simeq 0$, we can not distinguish a WD-PBH binary and a WD-WD binary.
Figure~\ref{fig: overlap} shows the distinguishability $x_\mathrm{dis}$ as a function of the WD's mass with $\mpbh=1.2M_\odot$ and $1.4M_\odot$.
Let us focus on the case of $\mpbh = 1.4M_\odot$ (thick black line in Fig.~\ref{fig: overlap}).
For $\mwd \gtrsim 0.55M_\odot$, the distinguishability $x_\mathrm{dis}$ of a WD-PBH binary and a WD-WD binary is almost unity. This is because the merger frequency of a WD-WD binary is lower than the initial frequency of a WD-PBH binary, as Fig.~\ref{fig: merger frequency zoom} shows. On the other hand, for $\mwd \lesssim 0.55 M_\odot$, the merger frequencies of a WD-WD binary and a WD-PBH binary are the same. This leads to no distinguishability in SNR.
Even in this case, WD-PBH binaries and WD-WD binaries could be distinguished once we go beyond the conservative approach. We can consider various effects related to the coalescence (e.g., tidal deformability). We will discuss it in more detail in the concluding section.

\section{Expected number of events}
\label{sec: expected number of events}

In this section, we estimate the merger rate of the WD-PBH binaries.
The expected number of events is given by
\begin{widetext}
    \begin{equation}
        N_\textrm{event}^\text{WD-PBH} = T_\textrm{obs} \int^{M_\textrm{WD,max}}_{M_\textrm{WD,min}} \frac{\dd{\mwd}}{\mwd}\ \int^{z_\textrm{up}}_0 \dd{z}\diff{V}{z} \int^{M_\textrm{h,max}}_{M_\textrm{h,min}} \dd{M_\textrm{h}} \diff{n_\textrm{halo}}{M_\textrm{h}} \dv{R_\textrm{WD-PBH}}{\ln\mwd}\,,
    \end{equation}
\end{widetext}
with the comoving volume $V(z)$ within the redshift $z$, the halo mass function $\displaystyle \dv{n_\textrm{halo}}{M_\textrm{h}}$, and the merger rate density per halo per logarithmic \ac{WD} mass $\displaystyle \dv{R_\textrm{WD-PBH}}{\ln \mwd}$.
We below neglect the redshift dependence of the halo mass function and the merger rate and use the current values for simplicity.
We set the minimum and maximum masses of halos at $M_{\uh,\umin}=400M_\odot$ and $M_{\uh,\umax}=10^{16} M_\odot$, respectively.
We have confirmed that the event rate of WD-PBH mergers is not very sensitive to the choice of $M_\mathrm{h,min}$ and $M_\mathrm{h,max}$.
\begin{figure}
    \centering
    \includegraphics[width=8cm]{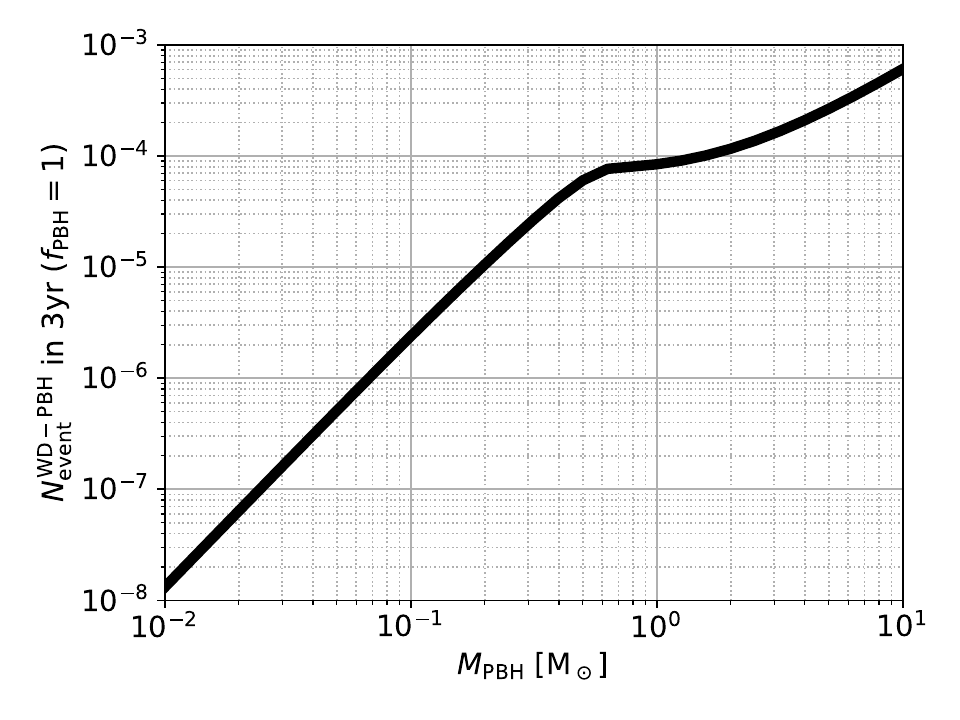}
    \caption{Expected number of WD-PBH mergers detected by three-year observation of DECIGO.}
    \label{fig: event number}
\end{figure}
The detailed description of each term in the integrand is given in Appendix~\ref{appendix}. Figure~\ref{fig: event number} shows the resultant event number in three years as a function of the \ac{PBH} mass.
Note that we assume the fraction of \ac{PBH} to the dark matter, $\fpbh$, is unity though astrophysical observations have already put the constraint $\fpbh \lesssim \calO(10^{-1})$ for $\mpbh = 0.1M_\odot$. 
Figure~\ref{fig: event number} implies that the merger rate of WD-PBH binaries is too low to detect them by three years of observation of DECIGO.

The small event number is because of our conservative estimation.
The expected event number would be increased by several orders of magnitude, depending on the assumptions on the number densities, the velocity dispersions, etc., of the target objects.
In the context of the NS-PBH merger, several possibilities (e.g., formation in ultra-faint dwarf galaxies or globular clusters and \ac{PBH} clustering at their formation)
provoking such enhancement are presented~\cite{Tsai:2020hpi}. Before concluding this section, we briefly see how much enhancement is required (and allowed) to detect the WD-PBH binaries.
Let us parametrize the enhancement of the event number by~\cite{Tsai:2020hpi}
\begin{equation}
    \tilde{N}^\textrm{WD-PBH}_\textrm{event} = \alpha_\textrm{WD} \times \alpha_\textrm{PBH} \times f_\textrm{PBH} \times N^\textrm{WD-PBH}_\textrm{event}\,,
    \label{eq: enhanced Nevent}
\end{equation}
where $\alpha_\textrm{WD}$ and $\alpha_\textrm{PBH}$ are the enhancement factors induced by some mechanisms related to the \Acp{WD} and the \Acp{PBH}, respectively.

Even forgetting about the realistic astrophysical processes, $\alpha_\PBH$ cannot be enlarged arbitrarily because it would conflict with the current constraint on the \ac{PBH}-\ac{PBH} merger rate achieved by the ground-based \ac{GW} detectors.
As we explained in Sec.~\ref{sec: introduction}, PBH-PBH binaries are formed in two channels, and the early-time formation basically dominates the late-time formation. However, if the enhancement factor $\alpha_\mathrm{PBH}$ for \ac{WD}-\ac{PBH} is too large, it would also enhance the late-time \ac{PBH}-\ac{PBH} formation, making it dominant and leading to the violation of the current constraint on the event rate of subsolar mass \ac{PBH} binary coalescences.
To discuss it quantitatively, we calculate the event number of PBH-PBH mergers in the late-time formation scenario.
Because $\alpha_\PBH$ phenomenologically parametrizes the enhancement due to the \ac{PBH} properties, the \ac{PBH}-\ac{PBH} merger rate would also be enhanced in a similar way as
\begin{equation}
    \tilde{N}^\textrm{PBH-PBH}_\textrm{event} \sim (\alpha_\textrm{PBH} f_\textrm{PBH})^2 N^\textrm{PBH-PBH}_\textrm{event}
\end{equation}
where $N^\textrm{PBH-PBH}_\textrm{event}$ is the expected number of PBH-PBH mergers estimated with $f_\textrm{PBH} = 1$ and $\alpha_\textrm{PBH}=1$. So far, binary mergers with masses of less than $1M_\odot$ have not been detected.
Therefore, $\tilde{N}^\textrm{PBH-PBH}_\textrm{event}$ should be less than unity. For $\mpbh = 10^{-1} M_\odot$ as our interest, we find $N^\textrm{PBH-PBH}_\textrm{event} \sim 1 \times 10^{-5}$ (see Fig.~\ref{fig: event number of PBHPBH merger} in the appendix). Therefore, we get the upper bound
\begin{equation}
    \tilde{N}^\textrm{PBH-PBH}_\textrm{event} \lesssim 1
    \quad\Rightarrow\quad \alpha_\textrm{PBH} f_\textrm{PBH} \lesssim 3 \times 10^2\,.
\end{equation}
Substituting this into $\tilde{N}^\textrm{WD-PBH}_\textrm{event}$ and taking the value $N^\textrm{WD-PBH}_\textrm{event} = 2 \times 10^{-6}$, the required enhancement factor $\alpha_\textrm{WD}$ for the detectable event rate of \ac{WD}-\ac{PBH} binaries would be
\begin{equation}
    \tilde{N}^\textrm{WD-PBH}_\textrm{event} \gtrsim 1
    \quad\Rightarrow\quad \alpha_\textrm{WD} \gtrsim 2\times 10^3\,.
\end{equation}
If such an enhancement would be realized, one or more WD-PBH merger events could be detected by DECIGO without contradicting the current constraint on the \ac{PBH}-\ac{PBH} merger rate.

\section{Conclusions and future perspective}
\label{sec: conclusion}

We have investigated the detectability of \ac{WD} and subsolar mass \ac{PBH} binaries formed by the \ac{GW} bremsstrahlung.
To this end, in Sec.~\ref{sec: detectability}, we first investigated the maximum distance of the detectable binaries as the horizon distance assuming the simple waveform model given by Eqs.~(\ref{eq: inspiral signal 1})--(\ref{eq: inspiral signal 2}) and
the three-year observation of DECIGO with its design sensitivity. 
We have shown that the \Acp{GW} from WD-PBH binaries at $\sim \SI{1}{Gpc}$ are detectable when their mass satisfies $M_{\textrm{WD}}\gtrsim 0.8 M_{\odot}$ and $M_{\textrm{PBH}} \gtrsim 0.5 M_{\odot}$. Second, we discussed the distinguishability between WD-PBH and WD-WD mergers by estimating the merger frequency from the comparison of their tidal and contact radii. We found that WD-PBH mergers can be distinguished from other binaries just by looking at the merger frequency unless the \ac{WD} mass is less than $\sim 0.55M_\odot$. Finally, We evaluated the merger rate and the expected number of WD-PBH merger events in Sec.~\ref{sec: expected number of events}. We assumed the standard halo properties and simplified \ac{WD} distributions. The conservative expectation of the number of WD-PBH merger events is too small ($\sim \calO(10^{-6})$) to be detected for $\mpbh \sim 10^{-1}M_\odot$. 

In a specific situation such as clustering distribution or low velocity-dispersion of target objects, the expected number of WD-PBH merger events would be much enhanced. We should note that an arbitrarily large enhancement due to some mechanisms related to \acp{PBH} would contradict the current constraint on the PBH-PBH merger rate in the ground-based detectors because the same mechanisms would also promote the \ac{PBH}-\ac{PBH} merger.
We conclude that $\calO(10^3)$ enhancement due to WD-related mechanisms could enable the detection of $\calO(1)$ WD-PBH merger events by three-year DECIGO observation without violating the current \ac{PBH}-\ac{PBH} constraint.

One may include additional effects to improve the detectability and distinguishability of PBH-WD binaries. For example, the waveform modeling can be improved by, e.g., the tidal deformability of \Acp{WD}. As a binary becomes closer, the \ac{WD} is gradually deformed, resulting in the modification in the gravitational waveform. This modification is significant around the merger.
Another example is the effect of mass transfer. The WD-WD binary under the stable mass transfer is discussed in the literature~\cite{Paczynski:1967, Kremer:2017xrg}. It is known that there is the possibility that the binary separation becomes wider as the mass transfers from the smaller star to the heavier star (called \emph{outspiral}). The \Acp{GW} from such a binary are much different from that of the case without the mass transfer.
The synergy between \Acp{GW} and electromagnetic emissions can also be beneficial in identifying the source. WD-WD binary mergers are candidates of type Ia supernovae~\cite{Iben:1984iz, Webbink:1984}. WD-PBH binary mergers could emit electromagnetic emissions that are different from those of WD-WD binary mergers. On the analogy of NS-BH binaries, tidal disruption events can also be expected from the WD-PBH binary mergers.
Numerical simulations for these merger events would be interesting (see~\cite{Markin:2023fxx} for a numerical simulation of a binary coalescence of a \ac{NS} and a subsolar mass \ac{PBH}).
As well as concrete mechanisms for the merger rate enhancement discussed in Sec.~\ref{sec: expected number of events}, these points should be worth studying in future work.

\acknowledgements
We thank Hiroki Takeda and Tomoya Kinugawa for the beneficial discussions.
This work is supported by JSPS KAKENHI Grants No.~JP21K13918 (YT), JP20H01899, JP23H04502 and JP23K13099 (TSY), JP20K03968 (SY), and JST SPRING Grant No.~JPMJSP2125 (RI).

\appendix
\section{Details of merger rate calculation}
\label{appendix}

\subsection{Halo density profile}

\begin{figure} 
    \centering
    \includegraphics[width=8cm]{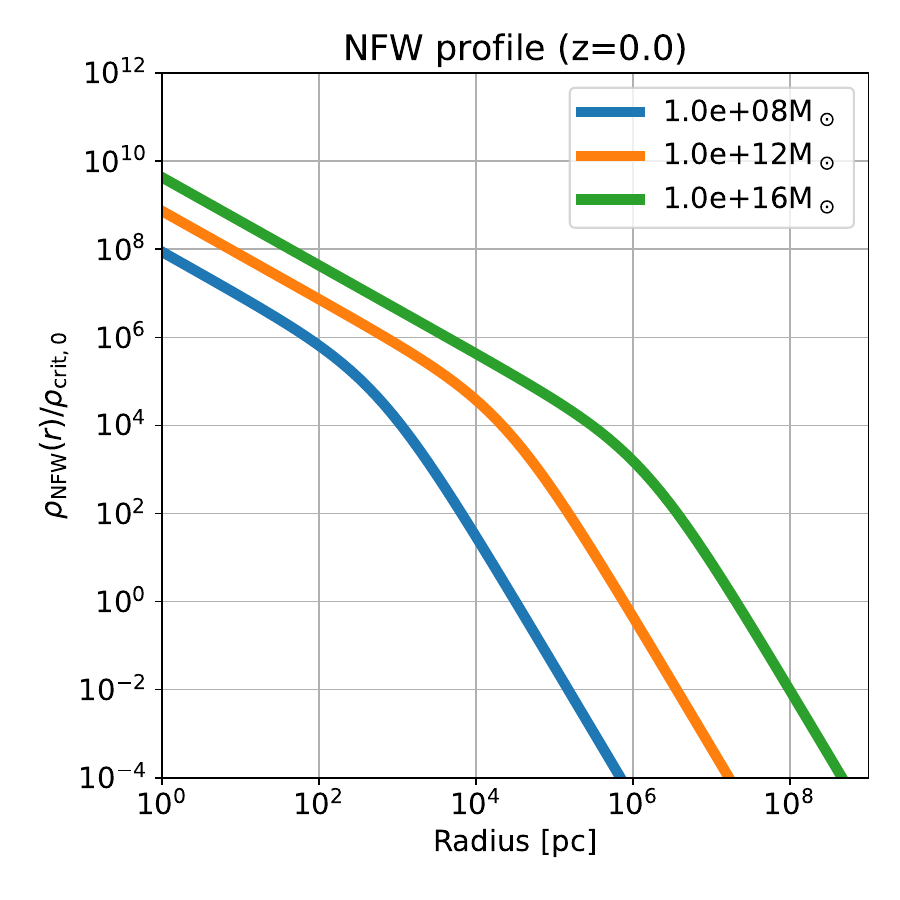}
    \caption{Examples of NFW profiles. We fix the redshift to $z=0$.}
    \label{fig:NFW profile}
\end{figure}

We employ the definition of the halo mass by the mass contained in a sphere where the averaged density is $200$ times larger than the critical density $\rho_\textrm{crit}$. The virial radius $r_\textrm{vir}$ is defined by the radius of the sphere which is used in the definition of the halo mass. In other words, $r_\textrm{vir}$ satisfies 
\begin{equation}
    M_\textrm{h} = \frac{4\pi}{3} r^3_\textrm{vir} \times200\rho_\textrm{crit} = 4\pi \int_0^{r_\textrm{vir}} \dd r\ r^2 \rho_\textrm{DM}(r)\,,
    \label{eq:virial radius}
\end{equation}
where $\rho_\textrm{DM}(r)$ is the dark matter density profile. We use the \ac{NFW} profile~\cite{Navarro:1995iw,Navarro:1996gj} for dark matter mass density profile (see Fig.~\ref{fig:NFW profile} for examples of this profile),
\begin{equation}
    \rho_\textrm{DM} (r)= \frac{\rho_s}{(r/r_\us) (1 + r/r_\us)^2}\,,
    \label{eq:NFW profile}
\end{equation}
where $\rho_\us$ and $r_\us$ characterize the density profile and are determined from the halo mass $M_\textrm{h}$ and the concentration factor $C \coloneqq r_\textrm{vir} / r_\textrm{s}$. The virial radius is given by the halo mass as shown in Eq.~\eqref{eq:virial radius}. So, we get the $r_\textrm{s}$ from $M_\textrm{h}$ and $C$. The density parameter $\rho_\textrm{s}$ is given by
\begin{equation}
    \frac{\rho_\us}{\rho_\textrm{crit}} = \frac{200}{3} \frac{C^3}{\ln(1+C) - C/(1+C)}\,.
\end{equation}

Ludlow~\etal~\cite{Ludlow:2016ifl} provides us with the fitting formula of the concentration parameter as a function of the halo mass and the redshift,
\begin{equation}
    C(\nu) = c_0 \left(\frac{\nu}{\nu_0} \right)^{-\gamma_1} \left[ 1 + \left(\frac{\nu}{\nu_0}\right)^{1/\beta} \right]^{-\beta(\gamma_2 - \gamma_1)}\,,
\end{equation}
where $\nu(z) \coloneqq \delta_\textrm{sc} / \sigma_\mathrm{m}(M,z)$ is the dimensionless peak height, $\delta_\textrm{sc} = 1.686$ is the spherical top-hat collapse threshold, and the parameters $\{c_0, \beta, \gamma_1, \gamma_2, \nu_0\}$ are
\begin{align}
    c_0 &= 3.395 \times x^{-0.215}\,,\\
    \beta &= 0.307 \times x^{0.540}\,,\\
    \gamma_1 &= 0.628 \times x^{-0.047}\,,\\
    \gamma_2 &= 0.317 \times x^{-0.893}\,,\\
    \nu_0 &= (4.135 - 0.564x - 0.210 x^2 \notag\\
    &\quad + 0.0557 x^3 - 0.00348 x^4) \times D(z)^{-1}\,,
\end{align}
with $x = 1+z$ and the linear growth factor $D(z)^{-1}$. These relation is valid for $0 \leq \log(1+z) \leq 1$ and $-8 \leq \log_{10} \bqty{ M / \pqty{h^{-1}M_\odot}} \leq 16.5$\,.
Here, $h$ is the Hubble parameter in the unit of $\SI{100}{km.s^{-1}.Mpc^{-1}}$.
The function $\sigma_\mathrm{m}(M, z)$ is the root mean square of the linear density field smoothed on the scale $R(M)=\bqty{3M/\pqty{4 \pi \bar{\rho}_\um}}^{1/3}$ and defined by
\begin{equation}\label{eq: variance sigma}
    \sigma_\mathrm{m}^2(M_\mathrm{h},z) = \frac{1}{2\pi^2} \int^\infty_0 \dd{k} k^2 P(k,z) \abs{\hat{W}(kR(M_\mathrm{h}))}^2\,,
\end{equation}
where $P(k)$ stands for the linear matter power spectrum and $\hat{W}(kR)$ is the Fourier transformation of the top-hat window function with radius $R$ defined in the real space.
$\sigma_\mathrm{m}$ is approximated by 
\begin{equation}
    \sigma_\mathrm{m}(M, z) = D(z)\frac{22.26 \xi^{0.292}}{1 + 1.53\xi^{0.275} + 3.36\xi^{0.198}}\,,
\end{equation}
where
\begin{equation}
    \xi \coloneqq \left(\frac{M_\textrm{h}}{10^{10} h^{-1} M_\odot} \right)^{-1}\,.
\end{equation}
The linear growth factor is approximated by
\begin{equation}
    D(z) = \frac{\Omega_\textrm{m}(z)}{\Omega_\textrm{m,0}} \frac{\Psi(0)}{\Psi(z)} (1+z)^{-1}\,,
\end{equation}
with
\beae{
    &\Psi(z) = \bmte{\Omega_\textrm{m}(z)^{4/7} - \Omega_\Lambda(z) \\ 
    	+ \left(1 + \frac{\Omega_\textrm{m}(z)}{2}\right) \left( 1 + \frac{\Omega_\Lambda(z)}{70} \right),} \\
    &\Omega_\Lambda(z) = \frac{\Omega_{\Lambda,0}}{\Omega_{\Lambda,0} + \Omega_{\textrm{m},0} (1+z)^3},
}
and $\Omega_\textrm{m}(z) = 1 - \Omega_\Lambda(z)$.

\subsection{Average of capture rate $\sigma \vrel$}

The cross-section $\sigma$ (not to confuse it with the perturbation amplitude $\sigma_\um$~\eqref{eq: variance sigma}) of a \ac{PBH} and a \ac{WD} with the relative velocity $v_\rel$ is
\begin{equation}
    \sigma = 2\pi \left(\frac{85\pi}{6\sqrt{2}} \right)^{2/7} \frac{G^2 (\mpbh + \mwd)^{10/7} \mwd^{2/7} \mpbh^{2/7}}{c^{10/7} v_\textrm{rel}^{18/7}}\,.
\end{equation}
The probability distribution of $\vrel$ would be approximated by the truncated Maxwell--Boltzmann distribution~\cite{Bird:2016dcv}:\footnote{Strictly speaking, Bird~\etal~\cite{Bird:2016dcv} employed this distribution for the \ac{PBH}-\ac{PBH} relative velocity.}
\bme{\label{eq: p vrel}
  p(\vrel ; \vvir, \vdm) \\
  = F_0 \qty[\ee^{-\vrel^2 / \vdm^2} - \ee^{-\vvir^2 / \vdm^2}]\Theta(\vvir-\vrel)\,,
}
where $\vvir$ is the virial velocity, $\vdm$ is the velocity dispersion of the dark matter (PBHs), and $F_0$ is the constant determined by the normalization
\begin{equation}
  1 = 4\pi \int^{\vvir}_0 \dd{v} v^2 p(v; \vvir, \vdm)\,.
\end{equation}
$\vdm$ and $\vvir$ can be related with halo parameters via Eq.~(6) of Ref.~\cite{Bird:2016dcv}: \begin{equation}\label{eq: vDM}
  \vdm = \sqrt{\frac{GM(r < r_\mathrm{max})}{r_\mathrm{max}}}
  = \frac{\vvir}{\sqrt{2}} \sqrt{\frac{C}{x_\mathrm{max}} \frac{g(x_\mathrm{max})}{g(C)}}\,,
\end{equation}
where $g(C) = \ln (1+C) - C/(1+C)$, and $x_\mathrm{max} = r_\mathrm{max} / r_s = 2.1626$ (equal to $C_m$ of Ref.~\cite{Bird:2016dcv}).
$M(r<r_\textrm{max})$ is the mass contained in the sphere of the radius $r_\textrm{max}$.
Using the NFW profile~\eqref{eq:NFW profile}, we get
\begin{align}
    M(r<r_\textrm{max}) &= 4\pi \int^{r_\textrm{max}}_0 \dd r\ r^2 \rho_\textrm{DM}(r) \notag\\
    &= 4\pi r_\textrm{s}^3 \rho_\textrm{s} \int^{x_\textrm{max}}_0 \dd x\ \frac{x}{(1+x)^2} \notag\\
    &= 4\pi r_\textrm{s}^3 \rho_\textrm{s} \left[ \ln(1+x_\textrm{max}) - \frac{x_\textrm{max}}{1 + x_\textrm{max}} \right]\,.
\end{align}
Once we get the parameters $\{r_\mathrm{s}, \rho_\mathrm{s}, C\}$, we obtain the velocity dispersion $\vdm$ from the first equality of Eq.~\eqref{eq: vDM} and the virial velocity $\vvir$ from the second equality.
The average of $\sigma \vrel$ is given by
\begin{equation}
    \langle \sigma \vrel \rangle = 4\pi \int^{\vvir}_0 \dd{v} v^2 p(v; \vvir, \vdm) \sigma v\,.
\end{equation}
Using Eqs.~\eqref{eq: p vrel} and~\eqref{eq: vDM}, we get
\begin{align}
    \langle \sigma v_\textrm{rel} \rangle &\simeq \mathsf{N} \times 2\pi \left(\frac{85\pi}{6\sqrt{2}} \right)^{2/7} \notag\\
    &\quad \times \frac{G^2 (\mpbh + \mwd)^{10/7} \mwd^{2/7} \mpbh^{2/7}}{c^{10/7} v_\textrm{DM}^{11/7}}\,,
\end{align}
where
\begin{equation}
    \mathsf{N} = \hat{F}_0 \int^{\hat{v}_\mathrm{vir}}_0 \dd{\hat{v}} \hat{v}^{3/7} \qty(e^{-\hat{v}^2} - e^{-\hat{v}_\mathrm{vir}^2})\,,
\end{equation}
with $\hat{v} = v / \vdm$, $\hat{v}_\mathrm{vir} = \vvir / \vdm$, and $\hat{F}_0 = 4\pi \vdm^3 F_0$.

\subsection{Merger rate in a particular halo}

The merger rate of a \ac{WD} with the mass of $M_\WD$ and a \ac{PBH} in a particular halo with the mass of $M_\textrm{h}$ is 
\begin{equation}
    \dv{R_\textrm{WD-PBH}}{\ln\mwd} = 4\pi \int^{r_\textrm{vir}}_0 \dd{r} r^2 \diff{n_\WD}{\ln\mwd} n_\PBH \langle \sigma v_\textrm{rel} \rangle\,,
\end{equation}
where $n_\WD$ and $n_\PBH$ are the number density profiles of \Acp{WD} and \Acp{PBH}, respectively.

For \Acp{PBH}, we simply assume that the density profile is proportional to the halo mass density,
\begin{equation}
    \rho_\textrm{PBH}(r) = \fpbh \rho_\textrm{DM}(r)\,,
\end{equation}
and the mass spectrum of \Acp{PBH} is monochromatic,
\begin{equation}
    n_\PBH(r) = \frac{\rho_\PBH(r)}{M_\PBH} = \fpbh \frac{\rho_\textrm{DM}(r)}{M_\PBH}\,.
\end{equation}

Sasaki~\etal~\cite{Sasaki:2021iuc} uses the exponential density profile for \Acp{NS}. We assume that the number density profile of the \Acp{WD} also is an exponential form, i.e.,
\begin{equation}
    {\dv{n_\WD}{\ln\mwd}}(r) = \mathcal{N}_\WD \ee^{-r/r_\WD}\,,
\end{equation}
where $\mathcal{N}_\WD$ is the normalization constant depending on the mass $\mwd$, and $r_\WD$ is the characteristic radius that is assumed to be independent of the mass. 
We assume~\cite{Sasaki:2021iuc}
\begin{equation}
    r_\WD = 0.1 r_\textrm{s}\,.
    \label{eq: def of r_wd}
\end{equation}
The normalization $\mathcal{N}_\WD$ is calculated as follows.
First, we start with the derivation of the galaxy's stellar mass.
We use the relation between the halo mass $M_\uh$ and the stellar mass $M_\ast$ of the galaxy. Behroozi~\etal~\cite{Behroozi_2010} gives the fitting function 
\bme{
    \log_{10} [M_\uh/M_\odot] = \log_{10} M_1 + \beta \log_{10} \left( \frac{M_\ast}{M_{\ast,0}}\right) \\
    + \frac{(M_\ast / M_{\ast,0})^{\delta}}{1 + (M_\ast / M_{\ast,0})^{-\gamma}} - \frac{1}{2}\,,
}
with
\begin{align}
    \log_{10} M_1 &= M_{1,0} + M_{1,a} (a - 1)\,,\\
    \log_{10} M_{\ast,0} &= M_{\ast,0,0} + M_{\ast,0,a} (a-1)\,,\\
    \beta &= \beta_0 + \beta_a (a-1)\,,\\
    \gamma &= \gamma_0 + \gamma_a (a-1)\,,\\
    \delta &= \delta_0 + \delta_a (a-1)\,,
\end{align}
where $a$ is the scale factor. The parameters are taken from Table 2 of Ref.~\cite{Behroozi_2010}: $M_{1,0} = 12.35$, $M_{1,a} = 0.28$, $M_{\ast,0,0} = 10.72$, $M_{\ast,0,a} = 0.55$, $\beta_0 = 0.44$, $\beta_a = 0.18$, $\gamma_0 = 1.56$, $\gamma_a = 2.51$, $\delta_0 = 0.57$, and $\delta_a = 0.17$. 
The galaxy's stellar mass determines the normalization constant of the mass spectrum of the main sequence stars through the condition,
\begin{align}
    M_\ast &= V \int^{M_\textrm{MS,max}}_{M_\textrm{MS,min}} \dd{M_\textrm{MS}}M_\textrm{MS} \diff{n}{M_\textrm{MS}} \notag\\
    &= V \int^{M_\textrm{MS,max}}_{M_\textrm{MS,min}} \dd{M_\textrm{MS}}\diff{n}{\ln M_\textrm{MS}}\,,
    \label{eq: M ast condition}
\end{align}
where $V$ is defined by
\begin{equation}
    V = \frac{4\pi}{3} r_\textrm{max}^3\,,
\end{equation}
and we set $M_\mathrm{MS, min}=0.01 M_\odot$ and $M_\mathrm{MS,max} = 100M_\odot$.
We use the following mass spectrum~\cite{Kroupa:2000iv},
\begin{align}
    &\frac{\mathrm{d} n}{\mathrm{d} \ln M_\mathrm{MS}} \notag\\
    &= \begin{cases}A_{\mathrm{BD}}\left(\frac{M_\MS}{0.08 M_{\odot}}\right)^{1-\alpha_{\mathrm{BD}}} & \left(0.01 \leq M_\MS / M_{\odot} \leq 0.08\right) \\
    A_{\mathrm{MS}}\left(\frac{M_\MS}{0.5 M_{\odot}}\right)^{1-\alpha_{\mathrm{MS} 1}} & \left(0.08 \leq M_\MS / M_{\odot} \leq 0.5\right) \\
    A_{\mathrm{MS}}\left(\frac{M_\MS}{0.5 M_{\odot}}\right)^{1-\alpha_{\mathrm{MS} 2}} & \left(M_\MS / M_{\odot} \geq 0.5\right)
    \end{cases}
\end{align}
where $\alpha_\textrm{BD} = 0.8$, $\alpha_\textrm{MS1} = 1.3$, and $\alpha_\textrm{MS2} = 2.0$~\cite{2017Natur.548..183M}. 
The constant $A_\textrm{MS}$ is determined by the continuity at $M_\MS=0.08M_\odot$, and $A_\textrm{BD}$ is chosen so that Eq.~\eqref{eq: M ast condition} holds.
We assume that the main sequence stars with masses of $1M_\odot \leq M_\MS \leq 8 M_\odot$ collapse to \Acp{WD}. Taking into account the mass loss of the stars, the mass of a \ac{WD} is estimated by the empirical relation~\cite{Williams:2008ms},
\begin{equation}
    \frac{\mwd}{M_\odot} = 0.339 + 0.129 \frac{M_\textrm{MS}}{M_\odot}\,,
\end{equation}
where $M_\textrm{MS}$ is a mass of the progenitor of the \ac{WD}. Using this relation, we can convert the mass spectrum of the main sequence stars to that of the \Acp{WD} which is denoted by
\begin{equation*}
    \diff{\bar{n}_\WD}{\ln \mwd}\,,
\end{equation*}
that has no radial dependence.
This number density is understood as the averaged one defined by
\begin{equation}
    \diff{\bar{n}_\WD}{\ln\mwd} = \frac{1}{V} \int_0^{r_\textrm{max}} \mathcal{N}_\WD \ee^{-r/r_\WD} \times4\pi r^2 \dd{r}.
\end{equation}
The normalization constant is hence determined by
\begin{align}
    \mathcal{N}_\WD &= \frac{V}{4\pi \int^{r_\textrm{max}}_0 \dd{r}r^2 \ee^{-r/r_\WD}} \diff{\bar{n}_\WD}{\ln\mwd} \notag\\
    &\simeq \frac{0.5 V}{4\pi r_\WD^3} \diff{\bar{n}_\WD}{\ln\mwd}\,.
\end{align}

\subsection{Merger rate per comoving volume and halo mass function}

\begin{figure}
    \centering
    \includegraphics[width=8cm]{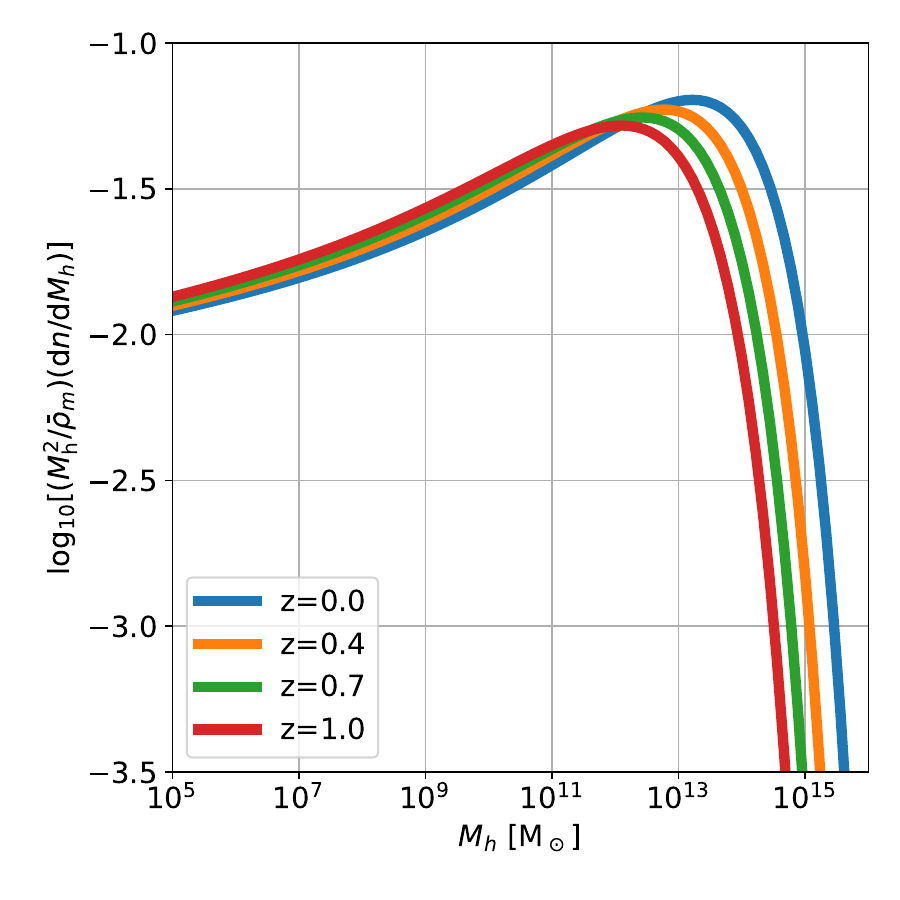}
    \caption{Halo mass function~\eqref{eq: halo mass function} in the form of $\log_{10}\pqty{\frac{M_\uh^2}{\bar{\rho}_\um}\dv{n_\halo}{M_\uh}}$.}
    \label{fig:halo mass function}
\end{figure}

Integrating the merger rate in a particular halo with a weight due to the halo mass function (i.e., the number of halos with the mass of $M_\textrm{h}$ per comoving volume), we get the merger rate of a particular mass of \ac{WD} per unit time per comoving volume by
\begin{equation}
    \int^{M_\textrm{h,max}}_{M_\textrm{h,min}} \dd{M_\textrm{h}} \diff{n_\textrm{halo}}{M_\textrm{h}} \dv{R_\textrm{WD-PBH}(M_\textrm{h})}{\ln \mwd}\,,
\end{equation}
where $\dv*{n_\textrm{halo}}{M_\textrm{h}}$ is the halo mass function. The fitting function of the halo mass function is given by Tinker~\etal~\cite{Tinker:2008ff}:
\begin{equation}\label{eq: halo mass function}
    \diff{n_\textrm{halo}}{M_\uh} = f(\sigma_\mathrm{m}) \frac{\bar{\rho}_\um}{M_\uh} \diff{\ln [\sigma_\mathrm{m}^{-1}]}{M_\uh}\,,
\end{equation}
with the matter density $\bar{\rho}_\um(z) = \Omega_\um(z) \rho_\textrm{crit}(z)$ and
\begin{equation}
    f(\sigma_\mathrm{m}) = A\left[ \left(\frac{\sigma_\mathrm{m}}{b}\right)^{-a} + 1 \right] \ee^{-c/\sigma_\mathrm{m}^2}\,.
\end{equation}
For $\{A, a,b, c\}$, we use the values for $\Delta =200$ (corresponding to the definition of the virial radius~\eqref{eq:virial radius}) in Table~2 in Ref.~\cite{Tinker:2008ff}, i.e.,
\begin{equation}
    A=0.186\,,\quad
    a=1.47\,,\quad
    b=2.57\,,\quad
    c=1.19\,.
\end{equation}
Figure~\ref{fig:halo mass function} shows the halo mass function with different redshifts.

\subsection{Rate of detectable mergers of \ac{PBH} and \ac{WD} with a particular mass $M_\WD$}

\begin{figure} 
    \centering
    \includegraphics[width=8cm]{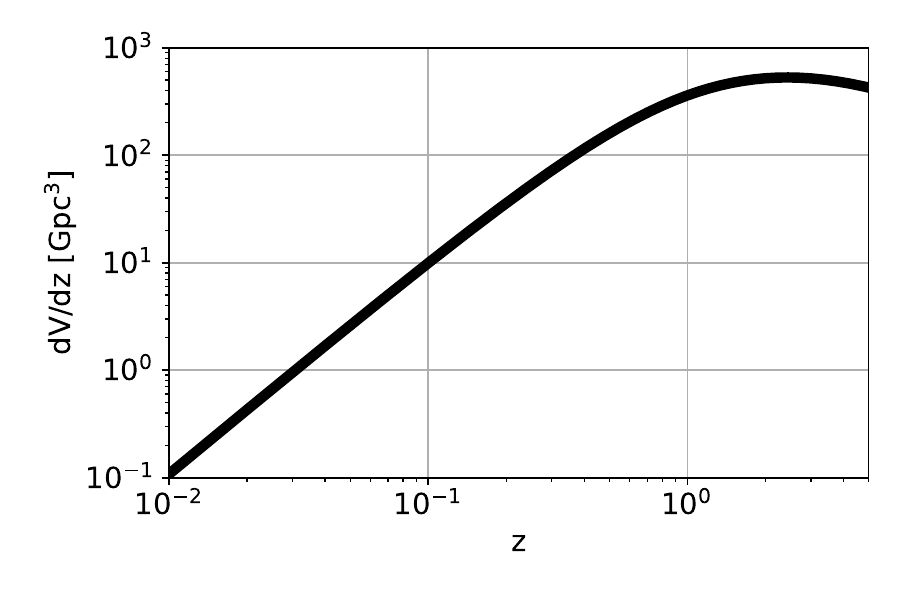}
    \caption{Unit comoving volume $\dv*{V}{z}$~\eqref{eq: unit comoving volume}.}
    \label{fig:unit comoving volume}
\end{figure}

The unit comoving volume is defined by
\begin{equation}
    \diff{V}{z} = \frac{c}{H(z)} 4\pi \left[\int^z_0 \frac{c \dd{z'}}{H(z')} \right]^2\,.
    \label{eq: unit comoving volume}
\end{equation}
Figure~\ref{fig:unit comoving volume} shows $\dv*{V}{z}$.

Using $\dv*{V}{z}$, we get the rate of the detectable mergers of \ac{PBH} and \ac{WD} with a particular mass $M_\WD$ by
\begin{equation}
    \int^{z_\textrm{up}}_0 \dd{z}\diff{V}{z}     \int^{M_\textrm{h,max}}_{M_\textrm{h,min}} \dd{M_\textrm{h}}\diff{n_\textrm{halo}}{M_\textrm{h}} \dv{R_\textrm{WD-PBH}(M_\textrm{h})}{\ln\mwd}\,.
\end{equation}
Here, $z_\textrm{up}$ is determined by the horizon distance,
\begin{equation}
    D_\textrm{horizon} = (1 + z_\textrm{up}) \int^{z_\textrm{up}}_0 \frac{c\dd{z'}}{H(z')}\,.
    \label{eq: WD-PBH rate with particular M_WD}
\end{equation}
The horizon distance depends on the \ac{WD} and \ac{PBH} masses and the detector's sensitivity, and so does $z_\textrm{up}$.

\subsection{Merger rate of PBH-PBH binaries formed in the late universe}

To calculate the merger rate of PBH-PBH binaries, we follow the same process as we did for PBH-WD binaries. The expected event number of PBH-PBH mergers is given by
\bme{
    N^\textrm{PBH-PBH}_\textrm{event} 
    \\
    = T_\textrm{obs} \int^{z_\textrm{up}}_0 \dd{z} \dv{V}{z} \int^{M_\textrm{h,max}}_{M_\textrm{h,min}} \dd{M_\textrm{h}} \dv{n_\textrm{halo}}{M_\textrm{h}} R_\textrm{PBH-PBH}\,.
}
The difference is the merger rate density per one halo.
Because we assume the monochromatic mass distribution for \Acp{PBH}, we do not carry out the integration over $\mpbh$.
The merger rate density per one halo $R_\mathrm{PBH-PBH}$ is given by
\begin{equation}
    R_\textrm{PBH-PBH} = 4\pi \int^{r_\textrm{vir}}_0 \dd{r} r^2 f_\textrm{PBH}^2 n_\textrm{DM}^2(r) \langle \sigma v_\textrm{rel} \rangle\,.
\end{equation}

\begin{figure}
    \centering
    \includegraphics[width=8cm]{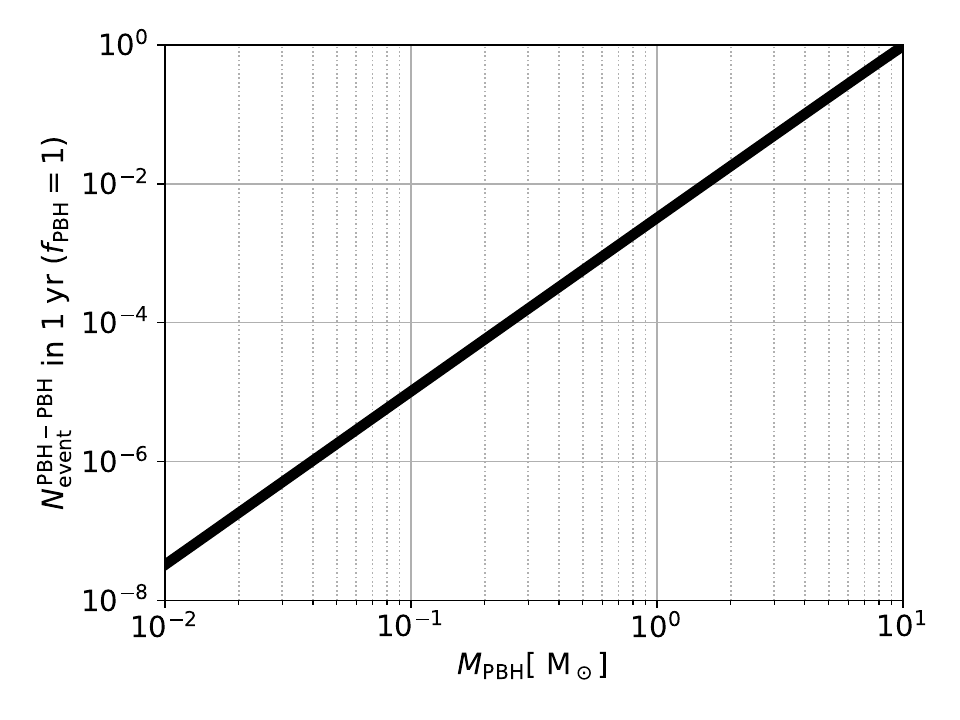}
    \caption{Expected number of PBH-PBH mergers detected by one-year observation by LIGO.}
    \label{fig: event number of PBHPBH merger}
\end{figure}

When we estimate $z_\mathrm{up}$, we use the PSD achieved in the first observation run of the LIGO-Virgo Collaboration,
\bme{
    S^\mathrm{LIGO}_\mathrm{n}(f) = \Bigg[
    \left(\frac{18\Hz}{0.1\Hz + f}\right)^4 \times 10^{-44} + 4.9 \times 10^{-47} \\
    + 1.6 \times 10^{-45} \left(\frac{f}{2000\Hz}\right)^2 \Bigg] \Hz^{-1}\,,
}
which is taken from Ref.~\cite{ligo_tutorial}. We assume the one-year observation. Figure~\ref{fig: event number of PBHPBH merger} shows the event rate of PBH-PBH binary mergers detected by one-year observation of the LIGO detector.
The result is consistent with preceding studies such as Bird~\etal~\cite{Bird:2016dcv}.

\bibliography{reference}

\begin{thebibliography}{60}%
\makeatletter
\providecommand \@ifxundefined [1]{%
 \@ifx{#1\undefined}
}%
\providecommand \@ifnum [1]{%
 \ifnum #1\expandafter \@firstoftwo
 \else \expandafter \@secondoftwo
 \fi
}%
\providecommand \@ifx [1]{%
 \ifx #1\expandafter \@firstoftwo
 \else \expandafter \@secondoftwo
 \fi
}%
\providecommand \natexlab [1]{#1}%
\providecommand \enquote  [1]{``#1''}%
\providecommand \bibnamefont  [1]{#1}%
\providecommand \bibfnamefont [1]{#1}%
\providecommand \citenamefont [1]{#1}%
\providecommand \href@noop [0]{\@secondoftwo}%
\providecommand \href [0]{\begingroup \@sanitize@url \@href}%
\providecommand \@href[1]{\@@startlink{#1}\@@href}%
\providecommand \@@href[1]{\endgroup#1\@@endlink}%
\providecommand \@sanitize@url [0]{\catcode `\\12\catcode `\$12\catcode `\&12\catcode `\#12\catcode `\^12\catcode `\_12\catcode `\%12\relax}%
\providecommand \@@startlink[1]{}%
\providecommand \@@endlink[0]{}%
\providecommand \url  [0]{\begingroup\@sanitize@url \@url }%
\providecommand \@url [1]{\endgroup\@href {#1}{\urlprefix }}%
\providecommand \urlprefix  [0]{URL }%
\providecommand \Eprint [0]{\href }%
\providecommand \doibase [0]{https://doi.org/}%
\providecommand \selectlanguage [0]{\@gobble}%
\providecommand \bibinfo  [0]{\@secondoftwo}%
\providecommand \bibfield  [0]{\@secondoftwo}%
\providecommand \translation [1]{[#1]}%
\providecommand \BibitemOpen [0]{}%
\providecommand \bibitemStop [0]{}%
\providecommand \bibitemNoStop [0]{.\EOS\space}%
\providecommand \EOS [0]{\spacefactor3000\relax}%
\providecommand \BibitemShut  [1]{\csname bibitem#1\endcsname}%
\let\auto@bib@innerbib\@empty
\bibitem [{\citenamefont {Zel'dovich}\ and\ \citenamefont {Novikov}(1967)}]{Zeldovich:1967lct}%
  \BibitemOpen
  \bibfield  {author} {\bibinfo {author} {\bibfnamefont {Y.~B.}\ \bibnamefont {Zel'dovich}}\ and\ \bibinfo {author} {\bibfnamefont {I.~D.}\ \bibnamefont {Novikov}},\ }\bibfield  {title} {\bibinfo {title} {{The Hypothesis of Cores Retarded during Expansion and the Hot Cosmological Model}},\ }\href@noop {} {\bibfield  {journal} {\bibinfo  {journal} {Soviet Astron. AJ (Engl. Transl. ),}\ }\textbf {\bibinfo {volume} {10}},\ \bibinfo {pages} {602} (\bibinfo {year} {1967})}\BibitemShut {NoStop}%
\bibitem [{\citenamefont {Hawking}(1971)}]{Hawking1971}%
  \BibitemOpen
  \bibfield  {author} {\bibinfo {author} {\bibfnamefont {S.}~\bibnamefont {Hawking}},\ }\bibfield  {title} {\bibinfo {title} {{Gravitationally Collapsed Objects of Very Low Mass}},\ }\href {https://doi.org/10.1093/mnras/152.1.75} {\bibfield  {journal} {\bibinfo  {journal} {Monthly Notices of the Royal Astronomical Society}\ }\textbf {\bibinfo {volume} {152}},\ \bibinfo {pages} {75} (\bibinfo {year} {1971})},\ \Eprint {https://arxiv.org/abs/https://academic.oup.com/mnras/article-pdf/152/1/75/9360899/mnras152-0075.pdf} {https://academic.oup.com/mnras/article-pdf/152/1/75/9360899/mnras152-0075.pdf} \BibitemShut {NoStop}%
\bibitem [{\citenamefont {Carr}\ and\ \citenamefont {Hawking}(1974)}]{Carr:1974nx}%
  \BibitemOpen
  \bibfield  {author} {\bibinfo {author} {\bibfnamefont {B.~J.}\ \bibnamefont {Carr}}\ and\ \bibinfo {author} {\bibfnamefont {S.~W.}\ \bibnamefont {Hawking}},\ }\bibfield  {title} {\bibinfo {title} {{Black holes in the early Universe}},\ }\href {https://doi.org/10.1093/mnras/168.2.399} {\bibfield  {journal} {\bibinfo  {journal} {Mon. Not. Roy. Astron. Soc.}\ }\textbf {\bibinfo {volume} {168}},\ \bibinfo {pages} {399} (\bibinfo {year} {1974})}\BibitemShut {NoStop}%
\bibitem [{\citenamefont {Carr}(1975)}]{Carr:1975qj}%
  \BibitemOpen
  \bibfield  {author} {\bibinfo {author} {\bibfnamefont {B.~J.}\ \bibnamefont {Carr}},\ }\bibfield  {title} {\bibinfo {title} {{The Primordial black hole mass spectrum}},\ }\href {https://doi.org/10.1086/153853} {\bibfield  {journal} {\bibinfo  {journal} {Astrophys. J.}\ }\textbf {\bibinfo {volume} {201}},\ \bibinfo {pages} {1} (\bibinfo {year} {1975})}\BibitemShut {NoStop}%
\bibitem [{\citenamefont {Chapline}(1975)}]{Chapline:1975ojl}%
  \BibitemOpen
  \bibfield  {author} {\bibinfo {author} {\bibfnamefont {G.~F.}\ \bibnamefont {Chapline}},\ }\bibfield  {title} {\bibinfo {title} {{Cosmological effects of primordial black holes}},\ }\href {https://doi.org/10.1038/253251a0} {\bibfield  {journal} {\bibinfo  {journal} {Nature}\ }\textbf {\bibinfo {volume} {253}},\ \bibinfo {pages} {251} (\bibinfo {year} {1975})}\BibitemShut {NoStop}%
\bibitem [{\citenamefont {Bean}\ and\ \citenamefont {Magueijo}(2002)}]{Bean:2002kx}%
  \BibitemOpen
  \bibfield  {author} {\bibinfo {author} {\bibfnamefont {R.}~\bibnamefont {Bean}}\ and\ \bibinfo {author} {\bibfnamefont {J.}~\bibnamefont {Magueijo}},\ }\bibfield  {title} {\bibinfo {title} {{Could supermassive black holes be quintessential primordial black holes?}},\ }\href {https://doi.org/10.1103/PhysRevD.66.063505} {\bibfield  {journal} {\bibinfo  {journal} {Phys. Rev. D}\ }\textbf {\bibinfo {volume} {66}},\ \bibinfo {pages} {063505} (\bibinfo {year} {2002})},\ \Eprint {https://arxiv.org/abs/astro-ph/0204486} {arXiv:astro-ph/0204486} \BibitemShut {NoStop}%
\bibitem [{\citenamefont {Duechting}(2004)}]{Duechting:2004dk}%
  \BibitemOpen
  \bibfield  {author} {\bibinfo {author} {\bibfnamefont {N.}~\bibnamefont {Duechting}},\ }\bibfield  {title} {\bibinfo {title} {{Supermassive black holes from primordial black hole seeds}},\ }\href {https://doi.org/10.1103/PhysRevD.70.064015} {\bibfield  {journal} {\bibinfo  {journal} {Phys. Rev. D}\ }\textbf {\bibinfo {volume} {70}},\ \bibinfo {pages} {064015} (\bibinfo {year} {2004})},\ \Eprint {https://arxiv.org/abs/astro-ph/0406260} {arXiv:astro-ph/0406260} \BibitemShut {NoStop}%
\bibitem [{\citenamefont {Liu}\ and\ \citenamefont {Bromm}(2022)}]{Liu:2022bvr}%
  \BibitemOpen
  \bibfield  {author} {\bibinfo {author} {\bibfnamefont {B.}~\bibnamefont {Liu}}\ and\ \bibinfo {author} {\bibfnamefont {V.}~\bibnamefont {Bromm}},\ }\bibfield  {title} {\bibinfo {title} {{Accelerating Early Massive Galaxy Formation with Primordial Black Holes}},\ }\href {https://doi.org/10.3847/2041-8213/ac927f} {\bibfield  {journal} {\bibinfo  {journal} {Astrophys. J. Lett.}\ }\textbf {\bibinfo {volume} {937}},\ \bibinfo {pages} {L30} (\bibinfo {year} {2022})},\ \Eprint {https://arxiv.org/abs/2208.13178} {arXiv:2208.13178 [astro-ph.CO]} \BibitemShut {NoStop}%
\bibitem [{\citenamefont {H\"utsi}\ \emph {et~al.}(2023)\citenamefont {H\"utsi}, \citenamefont {Raidal}, \citenamefont {Urrutia}, \citenamefont {Vaskonen},\ and\ \citenamefont {Veerm\"ae}}]{Hutsi:2022fzw}%
  \BibitemOpen
  \bibfield  {author} {\bibinfo {author} {\bibfnamefont {G.}~\bibnamefont {H\"utsi}}, \bibinfo {author} {\bibfnamefont {M.}~\bibnamefont {Raidal}}, \bibinfo {author} {\bibfnamefont {J.}~\bibnamefont {Urrutia}}, \bibinfo {author} {\bibfnamefont {V.}~\bibnamefont {Vaskonen}},\ and\ \bibinfo {author} {\bibfnamefont {H.}~\bibnamefont {Veerm\"ae}},\ }\bibfield  {title} {\bibinfo {title} {{Did JWST observe imprints of axion miniclusters or primordial black holes?}},\ }\href {https://doi.org/10.1103/PhysRevD.107.043502} {\bibfield  {journal} {\bibinfo  {journal} {Phys. Rev. D}\ }\textbf {\bibinfo {volume} {107}},\ \bibinfo {pages} {043502} (\bibinfo {year} {2023})},\ \Eprint {https://arxiv.org/abs/2211.02651} {arXiv:2211.02651 [astro-ph.CO]} \BibitemShut {NoStop}%
\bibitem [{\citenamefont {Niikura}\ \emph {et~al.}(2019{\natexlab{a}})\citenamefont {Niikura}, \citenamefont {Takada}, \citenamefont {Yokoyama}, \citenamefont {Sumi},\ and\ \citenamefont {Masaki}}]{Niikura:2019kqi}%
  \BibitemOpen
  \bibfield  {author} {\bibinfo {author} {\bibfnamefont {H.}~\bibnamefont {Niikura}}, \bibinfo {author} {\bibfnamefont {M.}~\bibnamefont {Takada}}, \bibinfo {author} {\bibfnamefont {S.}~\bibnamefont {Yokoyama}}, \bibinfo {author} {\bibfnamefont {T.}~\bibnamefont {Sumi}},\ and\ \bibinfo {author} {\bibfnamefont {S.}~\bibnamefont {Masaki}},\ }\bibfield  {title} {\bibinfo {title} {{Constraints on Earth-mass primordial black holes from OGLE 5-year microlensing events}},\ }\href {https://doi.org/10.1103/PhysRevD.99.083503} {\bibfield  {journal} {\bibinfo  {journal} {Phys. Rev. D}\ }\textbf {\bibinfo {volume} {99}},\ \bibinfo {pages} {083503} (\bibinfo {year} {2019}{\natexlab{a}})},\ \Eprint {https://arxiv.org/abs/1901.07120} {arXiv:1901.07120 [astro-ph.CO]} \BibitemShut {NoStop}%
\bibitem [{\citenamefont {Scholtz}\ and\ \citenamefont {Unwin}(2020)}]{Scholtz:2019csj}%
  \BibitemOpen
  \bibfield  {author} {\bibinfo {author} {\bibfnamefont {J.}~\bibnamefont {Scholtz}}\ and\ \bibinfo {author} {\bibfnamefont {J.}~\bibnamefont {Unwin}},\ }\bibfield  {title} {\bibinfo {title} {{What if Planet 9 is a Primordial Black Hole?}},\ }\href {https://doi.org/10.1103/PhysRevLett.125.051103} {\bibfield  {journal} {\bibinfo  {journal} {Phys. Rev. Lett.}\ }\textbf {\bibinfo {volume} {125}},\ \bibinfo {pages} {051103} (\bibinfo {year} {2020})},\ \Eprint {https://arxiv.org/abs/1909.11090} {arXiv:1909.11090 [hep-ph]} \BibitemShut {NoStop}%
\bibitem [{\citenamefont {Witten}(2020)}]{Witten:2020ifl}%
  \BibitemOpen
  \bibfield  {author} {\bibinfo {author} {\bibfnamefont {E.}~\bibnamefont {Witten}},\ }\bibfield  {title} {\bibinfo {title} {{Searching for a Black Hole in the Outer Solar System}},\ }\Eprint {https://arxiv.org/abs/2004.14192} {arXiv:2004.14192 [astro-ph.EP]}  (\bibinfo {year} {2020})\BibitemShut {NoStop}%
\bibitem [{\citenamefont {Smirnov}\ \emph {et~al.}(2022)\citenamefont {Smirnov}, \citenamefont {Goobar}, \citenamefont {Linden},\ and\ \citenamefont {M\"ortsell}}]{Smirnov:2022zip}%
  \BibitemOpen
  \bibfield  {author} {\bibinfo {author} {\bibfnamefont {J.}~\bibnamefont {Smirnov}}, \bibinfo {author} {\bibfnamefont {A.}~\bibnamefont {Goobar}}, \bibinfo {author} {\bibfnamefont {T.}~\bibnamefont {Linden}},\ and\ \bibinfo {author} {\bibfnamefont {E.}~\bibnamefont {M\"ortsell}},\ }\bibfield  {title} {\bibinfo {title} {{White Dwarfs in Dwarf Spheroidal Galaxies: A New Class of Compact-Dark-Matter Detectors}},\ }\Eprint {https://arxiv.org/abs/2211.00013} {arXiv:2211.00013 [astro-ph.CO]}  (\bibinfo {year} {2022})\BibitemShut {NoStop}%
\bibitem [{\citenamefont {Escriv\`a}\ \emph {et~al.}(2022)\citenamefont {Escriv\`a}, \citenamefont {Kuhnel},\ and\ \citenamefont {Tada}}]{Escriva:2022duf}%
  \BibitemOpen
  \bibfield  {author} {\bibinfo {author} {\bibfnamefont {A.}~\bibnamefont {Escriv\`a}}, \bibinfo {author} {\bibfnamefont {F.}~\bibnamefont {Kuhnel}},\ and\ \bibinfo {author} {\bibfnamefont {Y.}~\bibnamefont {Tada}},\ }\bibfield  {title} {\bibinfo {title} {{Primordial Black Holes}},\ }\Eprint {https://arxiv.org/abs/2211.05767} {arXiv:2211.05767 [astro-ph.CO]}  (\bibinfo {year} {2022})\BibitemShut {NoStop}%
\bibitem [{\citenamefont {Carr}\ \emph {et~al.}(2023)\citenamefont {Carr}, \citenamefont {Clesse}, \citenamefont {Garcia-Bellido}, \citenamefont {Hawkins},\ and\ \citenamefont {Kuhnel}}]{Carr:2023tpt}%
  \BibitemOpen
  \bibfield  {author} {\bibinfo {author} {\bibfnamefont {B.}~\bibnamefont {Carr}}, \bibinfo {author} {\bibfnamefont {S.}~\bibnamefont {Clesse}}, \bibinfo {author} {\bibfnamefont {J.}~\bibnamefont {Garcia-Bellido}}, \bibinfo {author} {\bibfnamefont {M.}~\bibnamefont {Hawkins}},\ and\ \bibinfo {author} {\bibfnamefont {F.}~\bibnamefont {Kuhnel}},\ }\bibfield  {title} {\bibinfo {title} {{Observational Evidence for Primordial Black Holes: A Positivist Perspective}},\ }\Eprint {https://arxiv.org/abs/2306.03903} {arXiv:2306.03903 [astro-ph.CO]}  (\bibinfo {year} {2023})\BibitemShut {NoStop}%
\bibitem [{\citenamefont {Niikura}\ \emph {et~al.}(2019{\natexlab{b}})\citenamefont {Niikura} \emph {et~al.}}]{Niikura:2017zjd}%
  \BibitemOpen
  \bibfield  {author} {\bibinfo {author} {\bibfnamefont {H.}~\bibnamefont {Niikura}} \emph {et~al.},\ }\bibfield  {title} {\bibinfo {title} {{Microlensing constraints on primordial black holes with Subaru/HSC Andromeda observations}},\ }\href {https://doi.org/10.1038/s41550-019-0723-1} {\bibfield  {journal} {\bibinfo  {journal} {Nature Astron.}\ }\textbf {\bibinfo {volume} {3}},\ \bibinfo {pages} {524} (\bibinfo {year} {2019}{\natexlab{b}})},\ \Eprint {https://arxiv.org/abs/1701.02151} {arXiv:1701.02151 [astro-ph.CO]} \BibitemShut {NoStop}%
\bibitem [{\citenamefont {Carr}\ \emph {et~al.}(2021)\citenamefont {Carr}, \citenamefont {Kohri}, \citenamefont {Sendouda},\ and\ \citenamefont {Yokoyama}}]{Carr:2020gox}%
  \BibitemOpen
  \bibfield  {author} {\bibinfo {author} {\bibfnamefont {B.}~\bibnamefont {Carr}}, \bibinfo {author} {\bibfnamefont {K.}~\bibnamefont {Kohri}}, \bibinfo {author} {\bibfnamefont {Y.}~\bibnamefont {Sendouda}},\ and\ \bibinfo {author} {\bibfnamefont {J.}~\bibnamefont {Yokoyama}},\ }\bibfield  {title} {\bibinfo {title} {{Constraints on primordial black holes}},\ }\href {https://doi.org/10.1088/1361-6633/ac1e31} {\bibfield  {journal} {\bibinfo  {journal} {Rept. Prog. Phys.}\ }\textbf {\bibinfo {volume} {84}},\ \bibinfo {pages} {116902} (\bibinfo {year} {2021})},\ \Eprint {https://arxiv.org/abs/2002.12778} {arXiv:2002.12778 [astro-ph.CO]} \BibitemShut {NoStop}%
\bibitem [{\citenamefont {Nakamura}\ \emph {et~al.}(1997)\citenamefont {Nakamura}, \citenamefont {Sasaki}, \citenamefont {Tanaka},\ and\ \citenamefont {Thorne}}]{Nakamura:1997sm}%
  \BibitemOpen
  \bibfield  {author} {\bibinfo {author} {\bibfnamefont {T.}~\bibnamefont {Nakamura}}, \bibinfo {author} {\bibfnamefont {M.}~\bibnamefont {Sasaki}}, \bibinfo {author} {\bibfnamefont {T.}~\bibnamefont {Tanaka}},\ and\ \bibinfo {author} {\bibfnamefont {K.~S.}\ \bibnamefont {Thorne}},\ }\bibfield  {title} {\bibinfo {title} {{Gravitational waves from coalescing black hole MACHO binaries}},\ }\href {https://doi.org/10.1086/310886} {\bibfield  {journal} {\bibinfo  {journal} {Astrophys. J. Lett.}\ }\textbf {\bibinfo {volume} {487}},\ \bibinfo {pages} {L139} (\bibinfo {year} {1997})},\ \Eprint {https://arxiv.org/abs/astro-ph/9708060} {arXiv:astro-ph/9708060} \BibitemShut {NoStop}%
\bibitem [{\citenamefont {Ioka}\ \emph {et~al.}(1998)\citenamefont {Ioka}, \citenamefont {Chiba}, \citenamefont {Tanaka},\ and\ \citenamefont {Nakamura}}]{Ioka:1998nz}%
  \BibitemOpen
  \bibfield  {author} {\bibinfo {author} {\bibfnamefont {K.}~\bibnamefont {Ioka}}, \bibinfo {author} {\bibfnamefont {T.}~\bibnamefont {Chiba}}, \bibinfo {author} {\bibfnamefont {T.}~\bibnamefont {Tanaka}},\ and\ \bibinfo {author} {\bibfnamefont {T.}~\bibnamefont {Nakamura}},\ }\bibfield  {title} {\bibinfo {title} {{Black hole binary formation in the expanding universe: Three body problem approximation}},\ }\href {https://doi.org/10.1103/PhysRevD.58.063003} {\bibfield  {journal} {\bibinfo  {journal} {Phys. Rev. D}\ }\textbf {\bibinfo {volume} {58}},\ \bibinfo {pages} {063003} (\bibinfo {year} {1998})},\ \Eprint {https://arxiv.org/abs/astro-ph/9807018} {arXiv:astro-ph/9807018} \BibitemShut {NoStop}%
\bibitem [{\citenamefont {Ali-Ha\"\i{}moud}\ \emph {et~al.}(2017)\citenamefont {Ali-Ha\"\i{}moud}, \citenamefont {Kovetz},\ and\ \citenamefont {Kamionkowski}}]{Ali-Haimoud:2017rtz}%
  \BibitemOpen
  \bibfield  {author} {\bibinfo {author} {\bibfnamefont {Y.}~\bibnamefont {Ali-Ha\"\i{}moud}}, \bibinfo {author} {\bibfnamefont {E.~D.}\ \bibnamefont {Kovetz}},\ and\ \bibinfo {author} {\bibfnamefont {M.}~\bibnamefont {Kamionkowski}},\ }\bibfield  {title} {\bibinfo {title} {{Merger rate of primordial black-hole binaries}},\ }\href {https://doi.org/10.1103/PhysRevD.96.123523} {\bibfield  {journal} {\bibinfo  {journal} {Phys. Rev. D}\ }\textbf {\bibinfo {volume} {96}},\ \bibinfo {pages} {123523} (\bibinfo {year} {2017})},\ \Eprint {https://arxiv.org/abs/1709.06576} {arXiv:1709.06576 [astro-ph.CO]} \BibitemShut {NoStop}%
\bibitem [{\citenamefont {Abbott}\ \emph {et~al.}(2018)\citenamefont {Abbott} \emph {et~al.}}]{LIGOScientific:2018glc}%
  \BibitemOpen
  \bibfield  {author} {\bibinfo {author} {\bibfnamefont {B.~P.}\ \bibnamefont {Abbott}} \emph {et~al.} (\bibinfo {collaboration} {LIGO Scientific, Virgo}),\ }\bibfield  {title} {\bibinfo {title} {{Search for Subsolar-Mass Ultracompact Binaries in Advanced LIGO\textquoteright{}s First Observing Run}},\ }\href {https://doi.org/10.1103/PhysRevLett.121.231103} {\bibfield  {journal} {\bibinfo  {journal} {Phys. Rev. Lett.}\ }\textbf {\bibinfo {volume} {121}},\ \bibinfo {pages} {231103} (\bibinfo {year} {2018})},\ \Eprint {https://arxiv.org/abs/1808.04771} {arXiv:1808.04771 [astro-ph.CO]} \BibitemShut {NoStop}%
\bibitem [{\citenamefont {Abbott}\ \emph {et~al.}(2019)\citenamefont {Abbott} \emph {et~al.}}]{LIGOScientific:2019kan}%
  \BibitemOpen
  \bibfield  {author} {\bibinfo {author} {\bibfnamefont {B.~P.}\ \bibnamefont {Abbott}} \emph {et~al.} (\bibinfo {collaboration} {LIGO Scientific, Virgo}),\ }\bibfield  {title} {\bibinfo {title} {{Search for Subsolar Mass Ultracompact Binaries in Advanced LIGO\textquoteright{}s Second Observing Run}},\ }\href {https://doi.org/10.1103/PhysRevLett.123.161102} {\bibfield  {journal} {\bibinfo  {journal} {Phys. Rev. Lett.}\ }\textbf {\bibinfo {volume} {123}},\ \bibinfo {pages} {161102} (\bibinfo {year} {2019})},\ \Eprint {https://arxiv.org/abs/1904.08976} {arXiv:1904.08976 [astro-ph.CO]} \BibitemShut {NoStop}%
\bibitem [{\citenamefont {Abbott}\ \emph {et~al.}(2022{\natexlab{a}})\citenamefont {Abbott} \emph {et~al.}}]{LIGOScientific:2021job}%
  \BibitemOpen
  \bibfield  {author} {\bibinfo {author} {\bibfnamefont {R.}~\bibnamefont {Abbott}} \emph {et~al.} (\bibinfo {collaboration} {LIGO Scientific, VIRGO, KAGRA}),\ }\bibfield  {title} {\bibinfo {title} {{Search for Subsolar-Mass Binaries in the First Half of Advanced LIGO\textquoteright{}s and Advanced Virgo\textquoteright{}s Third Observing Run}},\ }\href {https://doi.org/10.1103/PhysRevLett.129.061104} {\bibfield  {journal} {\bibinfo  {journal} {Phys. Rev. Lett.}\ }\textbf {\bibinfo {volume} {129}},\ \bibinfo {pages} {061104} (\bibinfo {year} {2022}{\natexlab{a}})},\ \Eprint {https://arxiv.org/abs/2109.12197} {arXiv:2109.12197 [astro-ph.CO]} \BibitemShut {NoStop}%
\bibitem [{\citenamefont {Abbott}\ \emph {et~al.}(2022{\natexlab{b}})\citenamefont {Abbott} \emph {et~al.}}]{LIGOScientific:2022hai}%
  \BibitemOpen
  \bibfield  {author} {\bibinfo {author} {\bibfnamefont {R.}~\bibnamefont {Abbott}} \emph {et~al.} (\bibinfo {collaboration} {LIGO Scientific, VIRGO, KAGRA}),\ }\bibfield  {title} {\bibinfo {title} {{Search for subsolar-mass black hole binaries in the second part of Advanced LIGO's and Advanced Virgo's third observing run}},\ }\Eprint {https://arxiv.org/abs/2212.01477} {arXiv:2212.01477 [astro-ph.HE]}  (\bibinfo {year} {2022}{\natexlab{b}})\BibitemShut {NoStop}%
\bibitem [{\citenamefont {Phukon}\ \emph {et~al.}(2021)\citenamefont {Phukon}, \citenamefont {Baltus}, \citenamefont {Caudill}, \citenamefont {Clesse}, \citenamefont {Depasse}, \citenamefont {Fays}, \citenamefont {Fong}, \citenamefont {Kapadia}, \citenamefont {Magee},\ and\ \citenamefont {Tanasijczuk}}]{Phukon:2021cus}%
  \BibitemOpen
  \bibfield  {author} {\bibinfo {author} {\bibfnamefont {K.~S.}\ \bibnamefont {Phukon}}, \bibinfo {author} {\bibfnamefont {G.}~\bibnamefont {Baltus}}, \bibinfo {author} {\bibfnamefont {S.}~\bibnamefont {Caudill}}, \bibinfo {author} {\bibfnamefont {S.}~\bibnamefont {Clesse}}, \bibinfo {author} {\bibfnamefont {A.}~\bibnamefont {Depasse}}, \bibinfo {author} {\bibfnamefont {M.}~\bibnamefont {Fays}}, \bibinfo {author} {\bibfnamefont {H.}~\bibnamefont {Fong}}, \bibinfo {author} {\bibfnamefont {S.~J.}\ \bibnamefont {Kapadia}}, \bibinfo {author} {\bibfnamefont {R.}~\bibnamefont {Magee}},\ and\ \bibinfo {author} {\bibfnamefont {A.~J.}\ \bibnamefont {Tanasijczuk}},\ }\bibfield  {title} {\bibinfo {title} {{The hunt for sub-solar primordial black holes in low mass ratio binaries is open}},\ }\Eprint {https://arxiv.org/abs/2105.11449} {arXiv:2105.11449 [astro-ph.CO]}  (\bibinfo {year} {2021})\BibitemShut {NoStop}%
\bibitem [{\citenamefont {{Morr{\'a}s}}\ \emph {et~al.}(2023)\citenamefont {{Morr{\'a}s}}, \citenamefont {{Nu{\~n}o Siles}}, \citenamefont {{Garc{\'\i}a-Bellido}}, \citenamefont {{Ruiz Morales}}, \citenamefont {{Men{\'e}ndez-V{\'a}zquez}}, \citenamefont {{Karathanasis}}, \citenamefont {{Martinovic}}, \citenamefont {{Phukon}}, \citenamefont {{Clesse}}, \citenamefont {{Mart{\'\i}nez}},\ and\ \citenamefont {{Sakellariadou}}}]{Morras:2023jvb}%
  \BibitemOpen
  \bibfield  {author} {\bibinfo {author} {\bibfnamefont {G.}~\bibnamefont {{Morr{\'a}s}}}, \bibinfo {author} {\bibfnamefont {J.~F.}\ \bibnamefont {{Nu{\~n}o Siles}}}, \bibinfo {author} {\bibfnamefont {J.}~\bibnamefont {{Garc{\'\i}a-Bellido}}}, \bibinfo {author} {\bibfnamefont {E.}~\bibnamefont {{Ruiz Morales}}}, \bibinfo {author} {\bibfnamefont {A.}~\bibnamefont {{Men{\'e}ndez-V{\'a}zquez}}}, \bibinfo {author} {\bibfnamefont {C.}~\bibnamefont {{Karathanasis}}}, \bibinfo {author} {\bibfnamefont {K.}~\bibnamefont {{Martinovic}}}, \bibinfo {author} {\bibfnamefont {K.~S.}\ \bibnamefont {{Phukon}}}, \bibinfo {author} {\bibfnamefont {S.}~\bibnamefont {{Clesse}}}, \bibinfo {author} {\bibfnamefont {M.}~\bibnamefont {{Mart{\'\i}nez}}},\ and\ \bibinfo {author} {\bibfnamefont {M.}~\bibnamefont {{Sakellariadou}}},\ }\bibfield  {title} {\bibinfo {title} {{Analysis of a subsolar-mass compact binary candidate from the second observing run of Advanced LIGO}},\ }\href {https://doi.org/10.1016/j.dark.2023.101285} {\bibfield
  {journal} {\bibinfo  {journal} {Physics of the Dark Universe}\ }\textbf {\bibinfo {volume} {42}},\ \bibinfo {eid} {101285} (\bibinfo {year} {2023})},\ \Eprint {https://arxiv.org/abs/2301.11619} {arXiv:2301.11619 [gr-qc]} \BibitemShut {NoStop}%
\bibitem [{\citenamefont {Prunier}\ \emph {et~al.}(2023)\citenamefont {Prunier}, \citenamefont {Morr\'as}, \citenamefont {Siles}, \citenamefont {Clesse}, \citenamefont {Garc\'\i{}a-Bellido},\ and\ \citenamefont {Ruiz~Morales}}]{Prunier:2023cyv}%
  \BibitemOpen
  \bibfield  {author} {\bibinfo {author} {\bibfnamefont {M.}~\bibnamefont {Prunier}}, \bibinfo {author} {\bibfnamefont {G.}~\bibnamefont {Morr\'as}}, \bibinfo {author} {\bibfnamefont {J.~F. N.~n.}\ \bibnamefont {Siles}}, \bibinfo {author} {\bibfnamefont {S.}~\bibnamefont {Clesse}}, \bibinfo {author} {\bibfnamefont {J.}~\bibnamefont {Garc\'\i{}a-Bellido}},\ and\ \bibinfo {author} {\bibfnamefont {E.}~\bibnamefont {Ruiz~Morales}},\ }\bibfield  {title} {\bibinfo {title} {{Analysis of the subsolar-mass black hole candidate SSM200308 from the second part of the third observing run of Advanced LIGO-Virgo}},\ }\Eprint {https://arxiv.org/abs/2311.16085} {arXiv:2311.16085 [gr-qc]}  (\bibinfo {year} {2023})\BibitemShut {NoStop}%
\bibitem [{\citenamefont {Pujolas}\ \emph {et~al.}(2021)\citenamefont {Pujolas}, \citenamefont {Vaskonen},\ and\ \citenamefont {Veerm\"ae}}]{Pujolas:2021yaw}%
  \BibitemOpen
  \bibfield  {author} {\bibinfo {author} {\bibfnamefont {O.}~\bibnamefont {Pujolas}}, \bibinfo {author} {\bibfnamefont {V.}~\bibnamefont {Vaskonen}},\ and\ \bibinfo {author} {\bibfnamefont {H.}~\bibnamefont {Veerm\"ae}},\ }\bibfield  {title} {\bibinfo {title} {{Prospects for probing gravitational waves from primordial black hole binaries}},\ }\href {https://doi.org/10.1103/PhysRevD.104.083521} {\bibfield  {journal} {\bibinfo  {journal} {Phys. Rev. D}\ }\textbf {\bibinfo {volume} {104}},\ \bibinfo {pages} {083521} (\bibinfo {year} {2021})},\ \Eprint {https://arxiv.org/abs/2107.03379} {arXiv:2107.03379 [astro-ph.CO]} \BibitemShut {NoStop}%
\bibitem [{\citenamefont {Maggiore}\ \emph {et~al.}(2020)\citenamefont {Maggiore} \emph {et~al.}}]{Maggiore:2019uih}%
  \BibitemOpen
  \bibfield  {author} {\bibinfo {author} {\bibfnamefont {M.}~\bibnamefont {Maggiore}} \emph {et~al.},\ }\bibfield  {title} {\bibinfo {title} {{Science Case for the Einstein Telescope}},\ }\href {https://doi.org/10.1088/1475-7516/2020/03/050} {\bibfield  {journal} {\bibinfo  {journal} {JCAP}\ }\textbf {\bibinfo {volume} {03}},\ \bibinfo {pages} {050}},\ \Eprint {https://arxiv.org/abs/1912.02622} {arXiv:1912.02622 [astro-ph.CO]} \BibitemShut {NoStop}%
\bibitem [{\citenamefont {Amaro-Seoane}\ \emph {et~al.}(2017)\citenamefont {Amaro-Seoane} \emph {et~al.}}]{LISA:2017pwj}%
  \BibitemOpen
  \bibfield  {author} {\bibinfo {author} {\bibfnamefont {P.}~\bibnamefont {Amaro-Seoane}} \emph {et~al.} (\bibinfo {collaboration} {LISA}),\ }\bibfield  {title} {\bibinfo {title} {{Laser Interferometer Space Antenna}},\ }\Eprint {https://arxiv.org/abs/1702.00786} {arXiv:1702.00786 [astro-ph.IM]}  (\bibinfo {year} {2017})\BibitemShut {NoStop}%
\bibitem [{\citenamefont {Seto}\ \emph {et~al.}(2001)\citenamefont {Seto}, \citenamefont {Kawamura},\ and\ \citenamefont {Nakamura}}]{Seto:2001qf}%
  \BibitemOpen
  \bibfield  {author} {\bibinfo {author} {\bibfnamefont {N.}~\bibnamefont {Seto}}, \bibinfo {author} {\bibfnamefont {S.}~\bibnamefont {Kawamura}},\ and\ \bibinfo {author} {\bibfnamefont {T.}~\bibnamefont {Nakamura}},\ }\bibfield  {title} {\bibinfo {title} {{Possibility of direct measurement of the acceleration of the universe using 0.1-Hz band laser interferometer gravitational wave antenna in space}},\ }\href {https://doi.org/10.1103/PhysRevLett.87.221103} {\bibfield  {journal} {\bibinfo  {journal} {Phys. Rev. Lett.}\ }\textbf {\bibinfo {volume} {87}},\ \bibinfo {pages} {221103} (\bibinfo {year} {2001})},\ \Eprint {https://arxiv.org/abs/astro-ph/0108011} {arXiv:astro-ph/0108011} \BibitemShut {NoStop}%
\bibitem [{\citenamefont {Kawamura}\ \emph {et~al.}(2021)\citenamefont {Kawamura} \emph {et~al.}}]{Kawamura:2020pcg}%
  \BibitemOpen
  \bibfield  {author} {\bibinfo {author} {\bibfnamefont {S.}~\bibnamefont {Kawamura}} \emph {et~al.},\ }\bibfield  {title} {\bibinfo {title} {{Current status of space gravitational wave antenna DECIGO and B-DECIGO}},\ }\href {https://doi.org/10.1093/ptep/ptab019} {\bibfield  {journal} {\bibinfo  {journal} {PTEP}\ }\textbf {\bibinfo {volume} {2021}},\ \bibinfo {pages} {05A105} (\bibinfo {year} {2021})},\ \Eprint {https://arxiv.org/abs/2006.13545} {arXiv:2006.13545 [gr-qc]} \BibitemShut {NoStop}%
\bibitem [{\citenamefont {{Quinlan}}\ and\ \citenamefont {{Shapiro}}(1989)}]{QuinlanShapiro1989}%
  \BibitemOpen
  \bibfield  {author} {\bibinfo {author} {\bibfnamefont {G.~D.}\ \bibnamefont {{Quinlan}}}\ and\ \bibinfo {author} {\bibfnamefont {S.~L.}\ \bibnamefont {{Shapiro}}},\ }\bibfield  {title} {\bibinfo {title} {{Dynamical Evolution of Dense Clusters of Compact Stars}},\ }\href {https://doi.org/10.1086/167745} {\bibfield  {journal} {\bibinfo  {journal} {\apj}\ }\textbf {\bibinfo {volume} {343}},\ \bibinfo {pages} {725} (\bibinfo {year} {1989})}\BibitemShut {NoStop}%
\bibitem [{\citenamefont {Mouri}\ and\ \citenamefont {Taniguchi}(2002)}]{Mouri:2002mc}%
  \BibitemOpen
  \bibfield  {author} {\bibinfo {author} {\bibfnamefont {H.}~\bibnamefont {Mouri}}\ and\ \bibinfo {author} {\bibfnamefont {Y.}~\bibnamefont {Taniguchi}},\ }\bibfield  {title} {\bibinfo {title} {{Runaway merging of black holes: analytical constraint on the timescale}},\ }\href {https://doi.org/10.1086/339472} {\bibfield  {journal} {\bibinfo  {journal} {Astrophys. J. Lett.}\ }\textbf {\bibinfo {volume} {566}},\ \bibinfo {pages} {L17} (\bibinfo {year} {2002})},\ \Eprint {https://arxiv.org/abs/astro-ph/0201102} {arXiv:astro-ph/0201102} \BibitemShut {NoStop}%
\bibitem [{\citenamefont {Tsai}\ \emph {et~al.}(2021)\citenamefont {Tsai}, \citenamefont {Palmese}, \citenamefont {Profumo},\ and\ \citenamefont {Jeltema}}]{Tsai:2020hpi}%
  \BibitemOpen
  \bibfield  {author} {\bibinfo {author} {\bibfnamefont {Y.-D.}\ \bibnamefont {Tsai}}, \bibinfo {author} {\bibfnamefont {A.}~\bibnamefont {Palmese}}, \bibinfo {author} {\bibfnamefont {S.}~\bibnamefont {Profumo}},\ and\ \bibinfo {author} {\bibfnamefont {T.}~\bibnamefont {Jeltema}},\ }\bibfield  {title} {\bibinfo {title} {{Is GW170817 a Multimessenger Neutron Star-Primordial Black Hole Merger?}},\ }\href {https://doi.org/10.1088/1475-7516/2021/10/019} {\bibfield  {journal} {\bibinfo  {journal} {JCAP}\ }\textbf {\bibinfo {volume} {10}},\ \bibinfo {pages} {019}},\ \Eprint {https://arxiv.org/abs/2007.03686} {arXiv:2007.03686 [astro-ph.HE]} \BibitemShut {NoStop}%
\bibitem [{\citenamefont {Sasaki}\ \emph {et~al.}(2022)\citenamefont {Sasaki}, \citenamefont {Takhistov}, \citenamefont {Vardanyan},\ and\ \citenamefont {Zhang}}]{Sasaki:2021iuc}%
  \BibitemOpen
  \bibfield  {author} {\bibinfo {author} {\bibfnamefont {M.}~\bibnamefont {Sasaki}}, \bibinfo {author} {\bibfnamefont {V.}~\bibnamefont {Takhistov}}, \bibinfo {author} {\bibfnamefont {V.}~\bibnamefont {Vardanyan}},\ and\ \bibinfo {author} {\bibfnamefont {Y.-l.}\ \bibnamefont {Zhang}},\ }\bibfield  {title} {\bibinfo {title} {{Establishing the Nonprimordial Origin of Black Hole\textendash{}Neutron Star Mergers}},\ }\href {https://doi.org/10.3847/1538-4357/ac66da} {\bibfield  {journal} {\bibinfo  {journal} {Astrophys. J.}\ }\textbf {\bibinfo {volume} {931}},\ \bibinfo {pages} {2} (\bibinfo {year} {2022})},\ \Eprint {https://arxiv.org/abs/2110.09509} {arXiv:2110.09509 [astro-ph.CO]} \BibitemShut {NoStop}%
\bibitem [{\citenamefont {Cholis}\ \emph {et~al.}(2016)\citenamefont {Cholis}, \citenamefont {Kovetz}, \citenamefont {Ali-Ha\"\i{}moud}, \citenamefont {Bird}, \citenamefont {Kamionkowski}, \citenamefont {Mu\~noz},\ and\ \citenamefont {Raccanelli}}]{Cholis:2016kqi}%
  \BibitemOpen
  \bibfield  {author} {\bibinfo {author} {\bibfnamefont {I.}~\bibnamefont {Cholis}}, \bibinfo {author} {\bibfnamefont {E.~D.}\ \bibnamefont {Kovetz}}, \bibinfo {author} {\bibfnamefont {Y.}~\bibnamefont {Ali-Ha\"\i{}moud}}, \bibinfo {author} {\bibfnamefont {S.}~\bibnamefont {Bird}}, \bibinfo {author} {\bibfnamefont {M.}~\bibnamefont {Kamionkowski}}, \bibinfo {author} {\bibfnamefont {J.~B.}\ \bibnamefont {Mu\~noz}},\ and\ \bibinfo {author} {\bibfnamefont {A.}~\bibnamefont {Raccanelli}},\ }\bibfield  {title} {\bibinfo {title} {{Orbital eccentricities in primordial black hole binaries}},\ }\href {https://doi.org/10.1103/PhysRevD.94.084013} {\bibfield  {journal} {\bibinfo  {journal} {Phys. Rev. D}\ }\textbf {\bibinfo {volume} {94}},\ \bibinfo {pages} {084013} (\bibinfo {year} {2016})},\ \Eprint {https://arxiv.org/abs/1606.07437} {arXiv:1606.07437 [astro-ph.HE]} \BibitemShut {NoStop}%
\bibitem [{ast()}]{astropy}%
  \BibitemOpen
  \href@noop {} {}\bibinfo {howpublished} {\url{http://www.astropy.org}}\BibitemShut {NoStop}%
\bibitem [{\citenamefont {Aghanim}\ \emph {et~al.}(2020)\citenamefont {Aghanim} \emph {et~al.}}]{Planck:2018vyg}%
  \BibitemOpen
  \bibfield  {author} {\bibinfo {author} {\bibfnamefont {N.}~\bibnamefont {Aghanim}} \emph {et~al.} (\bibinfo {collaboration} {Planck}),\ }\bibfield  {title} {\bibinfo {title} {{Planck 2018 results. VI. Cosmological parameters}},\ }\href {https://doi.org/10.1051/0004-6361/201833910} {\bibfield  {journal} {\bibinfo  {journal} {Astron. Astrophys.}\ }\textbf {\bibinfo {volume} {641}},\ \bibinfo {pages} {A6} (\bibinfo {year} {2020})},\ \bibinfo {note} {[Erratum: Astron.Astrophys. 652, C4 (2021)]},\ \Eprint {https://arxiv.org/abs/1807.06209} {arXiv:1807.06209 [astro-ph.CO]} \BibitemShut {NoStop}%
\bibitem [{\citenamefont {Dalal}\ \emph {et~al.}(2006)\citenamefont {Dalal}, \citenamefont {Holz}, \citenamefont {Hughes},\ and\ \citenamefont {Jain}}]{Dalal:2006qt}%
  \BibitemOpen
  \bibfield  {author} {\bibinfo {author} {\bibfnamefont {N.}~\bibnamefont {Dalal}}, \bibinfo {author} {\bibfnamefont {D.~E.}\ \bibnamefont {Holz}}, \bibinfo {author} {\bibfnamefont {S.~A.}\ \bibnamefont {Hughes}},\ and\ \bibinfo {author} {\bibfnamefont {B.}~\bibnamefont {Jain}},\ }\bibfield  {title} {\bibinfo {title} {{Short grb and binary black hole standard sirens as a probe of dark energy}},\ }\href {https://doi.org/10.1103/PhysRevD.74.063006} {\bibfield  {journal} {\bibinfo  {journal} {Phys. Rev. D}\ }\textbf {\bibinfo {volume} {74}},\ \bibinfo {pages} {063006} (\bibinfo {year} {2006})},\ \Eprint {https://arxiv.org/abs/astro-ph/0601275} {arXiv:astro-ph/0601275} \BibitemShut {NoStop}%
\bibitem [{\citenamefont {Yagi}\ and\ \citenamefont {Seto}(2011)}]{Yagi:2011wg}%
  \BibitemOpen
  \bibfield  {author} {\bibinfo {author} {\bibfnamefont {K.}~\bibnamefont {Yagi}}\ and\ \bibinfo {author} {\bibfnamefont {N.}~\bibnamefont {Seto}},\ }\bibfield  {title} {\bibinfo {title} {{Detector configuration of DECIGO/BBO and identification of cosmological neutron-star binaries}},\ }\href {https://doi.org/10.1103/PhysRevD.83.044011} {\bibfield  {journal} {\bibinfo  {journal} {Phys. Rev. D}\ }\textbf {\bibinfo {volume} {83}},\ \bibinfo {pages} {044011} (\bibinfo {year} {2011})},\ \bibinfo {note} {[Erratum: Phys.Rev.D 95, 109901 (2017)]},\ \Eprint {https://arxiv.org/abs/1101.3940} {arXiv:1101.3940 [astro-ph.CO]} \BibitemShut {NoStop}%
\bibitem [{\citenamefont {Magano}\ \emph {et~al.}(2017)\citenamefont {Magano}, \citenamefont {Vilas~Boas},\ and\ \citenamefont {Martins}}]{Magano:2017mqk}%
  \BibitemOpen
  \bibfield  {author} {\bibinfo {author} {\bibfnamefont {D.~M.~N.}\ \bibnamefont {Magano}}, \bibinfo {author} {\bibfnamefont {J.~M.~A.}\ \bibnamefont {Vilas~Boas}},\ and\ \bibinfo {author} {\bibfnamefont {C.~J. A.~P.}\ \bibnamefont {Martins}},\ }\bibfield  {title} {\bibinfo {title} {{Current and Future White Dwarf Mass-radius Constraints on Varying Fundamental Couplings and Unification Scenarios}},\ }\href {https://doi.org/10.1103/PhysRevD.96.083012} {\bibfield  {journal} {\bibinfo  {journal} {Phys. Rev. D}\ }\textbf {\bibinfo {volume} {96}},\ \bibinfo {pages} {083012} (\bibinfo {year} {2017})},\ \Eprint {https://arxiv.org/abs/1710.05828} {arXiv:1710.05828 [astro-ph.CO]} \BibitemShut {NoStop}%
\bibitem [{\citenamefont {Creighton}\ and\ \citenamefont {Anderson}(2011)}]{Creighton:2011zz}%
  \BibitemOpen
  \bibfield  {author} {\bibinfo {author} {\bibfnamefont {J.~D.~E.}\ \bibnamefont {Creighton}}\ and\ \bibinfo {author} {\bibfnamefont {W.~G.}\ \bibnamefont {Anderson}},\ }\href@noop {} {\emph {\bibinfo {title} {{Gravitational-wave physics and astronomy: An introduction to theory, experiment and data analysis}}}}\ (\bibinfo {year} {2011})\BibitemShut {NoStop}%
\bibitem [{\citenamefont {Maggiore}(2007)}]{Maggiore:2007ulw}%
  \BibitemOpen
  \bibfield  {author} {\bibinfo {author} {\bibfnamefont {M.}~\bibnamefont {Maggiore}},\ }\href@noop {} {\emph {\bibinfo {title} {{Gravitational Waves. Vol. 1: Theory and Experiments}}}},\ Oxford Master Series in Physics\ (\bibinfo  {publisher} {Oxford University Press},\ \bibinfo {year} {2007})\BibitemShut {NoStop}%
\bibitem [{\citenamefont {Jaranowski}\ and\ \citenamefont {Krolak}(2009)}]{Jaranowski:2009zz}%
  \BibitemOpen
  \bibfield  {author} {\bibinfo {author} {\bibfnamefont {P.}~\bibnamefont {Jaranowski}}\ and\ \bibinfo {author} {\bibfnamefont {A.}~\bibnamefont {Krolak}},\ }\href {https://doi.org/10.1017/CBO9780511605482} {\emph {\bibinfo {title} {{Analysis of gravitational-wave data}}}}\ (\bibinfo  {publisher} {Cambridge Univ. Press},\ \bibinfo {address} {Cambridge},\ \bibinfo {year} {2009})\BibitemShut {NoStop}%
\bibitem [{\citenamefont {{Paczy{\'n}ski}}(1967)}]{Paczynski:1967}%
  \BibitemOpen
  \bibfield  {author} {\bibinfo {author} {\bibfnamefont {B.}~\bibnamefont {{Paczy{\'n}ski}}},\ }\bibfield  {title} {\bibinfo {title} {{Gravitational Waves and the Evolution of Close Binaries}},\ }\href {https://ui.adsabs.harvard.edu/abs/1967AcA....17..287P} {\bibfield  {journal} {\bibinfo  {journal} {Acta Astron}\ }\textbf {\bibinfo {volume} {17}},\ \bibinfo {pages} {287} (\bibinfo {year} {1967})}\BibitemShut {NoStop}%
\bibitem [{\citenamefont {Kremer}\ \emph {et~al.}(2017)\citenamefont {Kremer}, \citenamefont {Breivik}, \citenamefont {Larson},\ and\ \citenamefont {Kalogera}}]{Kremer:2017xrg}%
  \BibitemOpen
  \bibfield  {author} {\bibinfo {author} {\bibfnamefont {K.}~\bibnamefont {Kremer}}, \bibinfo {author} {\bibfnamefont {K.}~\bibnamefont {Breivik}}, \bibinfo {author} {\bibfnamefont {S.~L.}\ \bibnamefont {Larson}},\ and\ \bibinfo {author} {\bibfnamefont {V.}~\bibnamefont {Kalogera}},\ }\bibfield  {title} {\bibinfo {title} {{Accreting Double white dwarf binaries: Implications for LISA}},\ }\href {https://doi.org/10.3847/1538-4357/aa8557} {\bibfield  {journal} {\bibinfo  {journal} {Astrophys. J.}\ }\textbf {\bibinfo {volume} {846}},\ \bibinfo {pages} {95} (\bibinfo {year} {2017})},\ \Eprint {https://arxiv.org/abs/1707.01104} {arXiv:1707.01104 [astro-ph.HE]} \BibitemShut {NoStop}%
\bibitem [{\citenamefont {Iben}\ and\ \citenamefont {Tutukov}(1984)}]{Iben:1984iz}%
  \BibitemOpen
  \bibfield  {author} {\bibinfo {author} {\bibfnamefont {I.}~\bibnamefont {Iben}, \bibfnamefont {Jr.}}\ and\ \bibinfo {author} {\bibfnamefont {A.~V.}\ \bibnamefont {Tutukov}},\ }\bibfield  {title} {\bibinfo {title} {{Supernovae of type I as end products of the evolution of binaries with components of moderate initial mass (M not greater than about 9 solar masses)}},\ }\href {https://doi.org/10.1086/190932} {\bibfield  {journal} {\bibinfo  {journal} {Astrophys. J. Suppl.}\ }\textbf {\bibinfo {volume} {54}},\ \bibinfo {pages} {335} (\bibinfo {year} {1984})}\BibitemShut {NoStop}%
\bibitem [{\citenamefont {{Webbink}}(1984)}]{Webbink:1984}%
  \BibitemOpen
  \bibfield  {author} {\bibinfo {author} {\bibfnamefont {R.~F.}\ \bibnamefont {{Webbink}}},\ }\bibfield  {title} {\bibinfo {title} {{Double white dwarfs as progenitors of R Coronae Borealis stars and type I supernovae.}},\ }\href {https://doi.org/10.1086/161701} {\bibfield  {journal} {\bibinfo  {journal} {\apj}\ }\textbf {\bibinfo {volume} {277}},\ \bibinfo {pages} {355} (\bibinfo {year} {1984})}\BibitemShut {NoStop}%
\bibitem [{\citenamefont {Markin}\ \emph {et~al.}(2023)\citenamefont {Markin}, \citenamefont {Neuweiler}, \citenamefont {Abac}, \citenamefont {Chaurasia}, \citenamefont {Ujevic}, \citenamefont {Bulla},\ and\ \citenamefont {Dietrich}}]{Markin:2023fxx}%
  \BibitemOpen
  \bibfield  {author} {\bibinfo {author} {\bibfnamefont {I.}~\bibnamefont {Markin}}, \bibinfo {author} {\bibfnamefont {A.}~\bibnamefont {Neuweiler}}, \bibinfo {author} {\bibfnamefont {A.}~\bibnamefont {Abac}}, \bibinfo {author} {\bibfnamefont {S.~V.}\ \bibnamefont {Chaurasia}}, \bibinfo {author} {\bibfnamefont {M.}~\bibnamefont {Ujevic}}, \bibinfo {author} {\bibfnamefont {M.}~\bibnamefont {Bulla}},\ and\ \bibinfo {author} {\bibfnamefont {T.}~\bibnamefont {Dietrich}},\ }\bibfield  {title} {\bibinfo {title} {{General-relativistic hydrodynamics simulation of a neutron star\textendash{}sub-solar-mass black hole merger}},\ }\href {https://doi.org/10.1103/PhysRevD.108.064025} {\bibfield  {journal} {\bibinfo  {journal} {Phys. Rev. D}\ }\textbf {\bibinfo {volume} {108}},\ \bibinfo {pages} {064025} (\bibinfo {year} {2023})},\ \Eprint {https://arxiv.org/abs/2304.11642} {arXiv:2304.11642 [gr-qc]} \BibitemShut {NoStop}%
\bibitem [{\citenamefont {Navarro}\ \emph {et~al.}(1996)\citenamefont {Navarro}, \citenamefont {Frenk},\ and\ \citenamefont {White}}]{Navarro:1995iw}%
  \BibitemOpen
  \bibfield  {author} {\bibinfo {author} {\bibfnamefont {J.~F.}\ \bibnamefont {Navarro}}, \bibinfo {author} {\bibfnamefont {C.~S.}\ \bibnamefont {Frenk}},\ and\ \bibinfo {author} {\bibfnamefont {S.~D.~M.}\ \bibnamefont {White}},\ }\bibfield  {title} {\bibinfo {title} {{The Structure of cold dark matter halos}},\ }\href {https://doi.org/10.1086/177173} {\bibfield  {journal} {\bibinfo  {journal} {Astrophys. J.}\ }\textbf {\bibinfo {volume} {462}},\ \bibinfo {pages} {563} (\bibinfo {year} {1996})},\ \Eprint {https://arxiv.org/abs/astro-ph/9508025} {arXiv:astro-ph/9508025} \BibitemShut {NoStop}%
\bibitem [{\citenamefont {Navarro}\ \emph {et~al.}(1997)\citenamefont {Navarro}, \citenamefont {Frenk},\ and\ \citenamefont {White}}]{Navarro:1996gj}%
  \BibitemOpen
  \bibfield  {author} {\bibinfo {author} {\bibfnamefont {J.~F.}\ \bibnamefont {Navarro}}, \bibinfo {author} {\bibfnamefont {C.~S.}\ \bibnamefont {Frenk}},\ and\ \bibinfo {author} {\bibfnamefont {S.~D.~M.}\ \bibnamefont {White}},\ }\bibfield  {title} {\bibinfo {title} {{A Universal density profile from hierarchical clustering}},\ }\href {https://doi.org/10.1086/304888} {\bibfield  {journal} {\bibinfo  {journal} {Astrophys. J.}\ }\textbf {\bibinfo {volume} {490}},\ \bibinfo {pages} {493} (\bibinfo {year} {1997})},\ \Eprint {https://arxiv.org/abs/astro-ph/9611107} {arXiv:astro-ph/9611107} \BibitemShut {NoStop}%
\bibitem [{\citenamefont {Ludlow}\ \emph {et~al.}(2016)\citenamefont {Ludlow}, \citenamefont {Bose}, \citenamefont {Angulo}, \citenamefont {Wang}, \citenamefont {Hellwing}, \citenamefont {Navarro}, \citenamefont {Cole},\ and\ \citenamefont {Frenk}}]{Ludlow:2016ifl}%
  \BibitemOpen
  \bibfield  {author} {\bibinfo {author} {\bibfnamefont {A.~D.}\ \bibnamefont {Ludlow}}, \bibinfo {author} {\bibfnamefont {S.}~\bibnamefont {Bose}}, \bibinfo {author} {\bibfnamefont {R.~E.}\ \bibnamefont {Angulo}}, \bibinfo {author} {\bibfnamefont {L.}~\bibnamefont {Wang}}, \bibinfo {author} {\bibfnamefont {W.~A.}\ \bibnamefont {Hellwing}}, \bibinfo {author} {\bibfnamefont {J.~F.}\ \bibnamefont {Navarro}}, \bibinfo {author} {\bibfnamefont {S.}~\bibnamefont {Cole}},\ and\ \bibinfo {author} {\bibfnamefont {C.~S.}\ \bibnamefont {Frenk}},\ }\bibfield  {title} {\bibinfo {title} {{The mass\textendash{}concentration\textendash{}redshift relation of cold and warm dark matter haloes}},\ }\href {https://doi.org/10.1093/mnras/stw1046} {\bibfield  {journal} {\bibinfo  {journal} {Mon. Not. Roy. Astron. Soc.}\ }\textbf {\bibinfo {volume} {460}},\ \bibinfo {pages} {1214} (\bibinfo {year} {2016})},\ \Eprint {https://arxiv.org/abs/1601.02624} {arXiv:1601.02624 [astro-ph.CO]} \BibitemShut {NoStop}%
\bibitem [{\citenamefont {Bird}\ \emph {et~al.}(2016)\citenamefont {Bird}, \citenamefont {Cholis}, \citenamefont {Mu\~noz}, \citenamefont {Ali-Ha\"\i{}moud}, \citenamefont {Kamionkowski}, \citenamefont {Kovetz}, \citenamefont {Raccanelli},\ and\ \citenamefont {Riess}}]{Bird:2016dcv}%
  \BibitemOpen
  \bibfield  {author} {\bibinfo {author} {\bibfnamefont {S.}~\bibnamefont {Bird}}, \bibinfo {author} {\bibfnamefont {I.}~\bibnamefont {Cholis}}, \bibinfo {author} {\bibfnamefont {J.~B.}\ \bibnamefont {Mu\~noz}}, \bibinfo {author} {\bibfnamefont {Y.}~\bibnamefont {Ali-Ha\"\i{}moud}}, \bibinfo {author} {\bibfnamefont {M.}~\bibnamefont {Kamionkowski}}, \bibinfo {author} {\bibfnamefont {E.~D.}\ \bibnamefont {Kovetz}}, \bibinfo {author} {\bibfnamefont {A.}~\bibnamefont {Raccanelli}},\ and\ \bibinfo {author} {\bibfnamefont {A.~G.}\ \bibnamefont {Riess}},\ }\bibfield  {title} {\bibinfo {title} {{Did LIGO detect dark matter?}},\ }\href {https://doi.org/10.1103/PhysRevLett.116.201301} {\bibfield  {journal} {\bibinfo  {journal} {Phys. Rev. Lett.}\ }\textbf {\bibinfo {volume} {116}},\ \bibinfo {pages} {201301} (\bibinfo {year} {2016})},\ \Eprint {https://arxiv.org/abs/1603.00464} {arXiv:1603.00464 [astro-ph.CO]} \BibitemShut {NoStop}%
\bibitem [{\citenamefont {Behroozi}\ \emph {et~al.}(2010)\citenamefont {Behroozi}, \citenamefont {Conroy},\ and\ \citenamefont {Wechsler}}]{Behroozi_2010}%
  \BibitemOpen
  \bibfield  {author} {\bibinfo {author} {\bibfnamefont {P.~S.}\ \bibnamefont {Behroozi}}, \bibinfo {author} {\bibfnamefont {C.}~\bibnamefont {Conroy}},\ and\ \bibinfo {author} {\bibfnamefont {R.~H.}\ \bibnamefont {Wechsler}},\ }\bibfield  {title} {\bibinfo {title} {{A Comprehensive Analysis of Uncertainties Affecting the Stellar Mass-Halo Mass Relation for 0 \ensuremath{<} z \ensuremath{<} 4}},\ }\href {https://doi.org/10.1088/0004-637X/717/1/379} {\bibfield  {journal} {\bibinfo  {journal} {Astrophys. J.}\ }\textbf {\bibinfo {volume} {717}},\ \bibinfo {pages} {379} (\bibinfo {year} {2010})},\ \Eprint {https://arxiv.org/abs/1001.0015} {arXiv:1001.0015 [astro-ph.CO]} \BibitemShut {NoStop}%
\bibitem [{\citenamefont {Kroupa}(2001)}]{Kroupa:2000iv}%
  \BibitemOpen
  \bibfield  {author} {\bibinfo {author} {\bibfnamefont {P.}~\bibnamefont {Kroupa}},\ }\bibfield  {title} {\bibinfo {title} {{On the variation of the initial mass function}},\ }\href {https://doi.org/10.1046/j.1365-8711.2001.04022.x} {\bibfield  {journal} {\bibinfo  {journal} {Mon. Not. Roy. Astron. Soc.}\ }\textbf {\bibinfo {volume} {322}},\ \bibinfo {pages} {231} (\bibinfo {year} {2001})},\ \Eprint {https://arxiv.org/abs/astro-ph/0009005} {arXiv:astro-ph/0009005} \BibitemShut {NoStop}%
\bibitem [{\citenamefont {{Mr{\'o}z}}\ \emph {et~al.}(2017)\citenamefont {{Mr{\'o}z}}, \citenamefont {{Udalski}}, \citenamefont {{Skowron}}, \citenamefont {{Poleski}}, \citenamefont {{Koz{\l}owski}}, \citenamefont {{Szyma{\'n}ski}}, \citenamefont {{Soszy{\'n}ski}}, \citenamefont {{Wyrzykowski}}, \citenamefont {{Pietrukowicz}}, \citenamefont {{Ulaczyk}}, \citenamefont {{Skowron}},\ and\ \citenamefont {{Pawlak}}}]{2017Natur.548..183M}%
  \BibitemOpen
  \bibfield  {author} {\bibinfo {author} {\bibfnamefont {P.}~\bibnamefont {{Mr{\'o}z}}}, \bibinfo {author} {\bibfnamefont {A.}~\bibnamefont {{Udalski}}}, \bibinfo {author} {\bibfnamefont {J.}~\bibnamefont {{Skowron}}}, \bibinfo {author} {\bibfnamefont {R.}~\bibnamefont {{Poleski}}}, \bibinfo {author} {\bibfnamefont {S.}~\bibnamefont {{Koz{\l}owski}}}, \bibinfo {author} {\bibfnamefont {M.~K.}\ \bibnamefont {{Szyma{\'n}ski}}}, \bibinfo {author} {\bibfnamefont {I.}~\bibnamefont {{Soszy{\'n}ski}}}, \bibinfo {author} {\bibfnamefont {{\L}.}~\bibnamefont {{Wyrzykowski}}}, \bibinfo {author} {\bibfnamefont {P.}~\bibnamefont {{Pietrukowicz}}}, \bibinfo {author} {\bibfnamefont {K.}~\bibnamefont {{Ulaczyk}}}, \bibinfo {author} {\bibfnamefont {D.}~\bibnamefont {{Skowron}}},\ and\ \bibinfo {author} {\bibfnamefont {M.}~\bibnamefont {{Pawlak}}},\ }\bibfield  {title} {\bibinfo {title} {{No large population of unbound or wide-orbit Jupiter-mass planets}},\ }\href {https://doi.org/10.1038/nature23276} {\bibfield  {journal}
  {\bibinfo  {journal} {\nat}\ }\textbf {\bibinfo {volume} {548}},\ \bibinfo {pages} {183} (\bibinfo {year} {2017})},\ \Eprint {https://arxiv.org/abs/1707.07634} {arXiv:1707.07634 [astro-ph.EP]} \BibitemShut {NoStop}%
\bibitem [{\citenamefont {Williams}\ \emph {et~al.}(2009)\citenamefont {Williams}, \citenamefont {Bolte},\ and\ \citenamefont {Koester}}]{Williams:2008ms}%
  \BibitemOpen
  \bibfield  {author} {\bibinfo {author} {\bibfnamefont {K.~A.}\ \bibnamefont {Williams}}, \bibinfo {author} {\bibfnamefont {M.}~\bibnamefont {Bolte}},\ and\ \bibinfo {author} {\bibfnamefont {D.}~\bibnamefont {Koester}},\ }\bibfield  {title} {\bibinfo {title} {{Probing The Lower Mass Limit for Supernova Progenitors and the High-Mass End of the Initial-Final Mass Relation from White Dwarfs in the Open Cluster M35 (NGC 2168)}},\ }\href {https://doi.org/10.1088/0004-637X/693/1/355} {\bibfield  {journal} {\bibinfo  {journal} {Astrophys. J.}\ }\textbf {\bibinfo {volume} {693}},\ \bibinfo {pages} {355} (\bibinfo {year} {2009})},\ \Eprint {https://arxiv.org/abs/0811.1577} {arXiv:0811.1577 [astro-ph]} \BibitemShut {NoStop}%
\bibitem [{\citenamefont {Tinker}\ \emph {et~al.}(2008)\citenamefont {Tinker}, \citenamefont {Kravtsov}, \citenamefont {Klypin}, \citenamefont {Abazajian}, \citenamefont {Warren}, \citenamefont {Yepes}, \citenamefont {Gottlober},\ and\ \citenamefont {Holz}}]{Tinker:2008ff}%
  \BibitemOpen
  \bibfield  {author} {\bibinfo {author} {\bibfnamefont {J.~L.}\ \bibnamefont {Tinker}}, \bibinfo {author} {\bibfnamefont {A.~V.}\ \bibnamefont {Kravtsov}}, \bibinfo {author} {\bibfnamefont {A.}~\bibnamefont {Klypin}}, \bibinfo {author} {\bibfnamefont {K.}~\bibnamefont {Abazajian}}, \bibinfo {author} {\bibfnamefont {M.~S.}\ \bibnamefont {Warren}}, \bibinfo {author} {\bibfnamefont {G.}~\bibnamefont {Yepes}}, \bibinfo {author} {\bibfnamefont {S.}~\bibnamefont {Gottlober}},\ and\ \bibinfo {author} {\bibfnamefont {D.~E.}\ \bibnamefont {Holz}},\ }\bibfield  {title} {\bibinfo {title} {{Toward a halo mass function for precision cosmology: The Limits of universality}},\ }\href {https://doi.org/10.1086/591439} {\bibfield  {journal} {\bibinfo  {journal} {Astrophys. J.}\ }\textbf {\bibinfo {volume} {688}},\ \bibinfo {pages} {709} (\bibinfo {year} {2008})},\ \Eprint {https://arxiv.org/abs/0803.2706} {arXiv:0803.2706 [astro-ph]} \BibitemShut {NoStop}%
\bibitem [{lig()}]{ligo_tutorial}%
  \BibitemOpen
  \href@noop {} {\bibinfo {title} {{Binary black hole signals in LIGO open data}}},\ \bibinfo {howpublished} {\url{https://github.com/losc-tutorial/LOSC_Event_tutorial/blob/master/LOSC_Event_tutorial.ipynb}}\BibitemShut {NoStop}%
\end{thebibliography}%

\end{document}